\documentclass[10pt]{article}
\usepackage{amssymb}
\usepackage{amsmath}
\usepackage{amsthm}
\usepackage{color}
\usepackage{comment}
\usepackage{geometry}		
\usepackage{graphicx}
\usepackage{siunitx}

\let\originalleft\left
\let\originalright\right
\renewcommand{\left}{\mathopen{}\mathclose\bgroup\originalleft}
\renewcommand{\right}{\aftergroup\egroup\originalright}

\usepackage{tabularx,booktabs}
\usepackage{caption}
\usepackage{subcaption}

\usepackage{hyperref}
\hypersetup{
    colorlinks=true,
    linkcolor=blue,
    urlcolor=blue,
    citecolor=blue
}
\newcommand{\myEdit}[1]{{\color{black} #1}}

\usepackage{tikz}
\usepackage{array}
\usepackage{booktabs}
\usepackage{epstopdf}

\usepackage{fancyhdr}

\usepackage{mathtools}
\usepackage{tensor}

\newtheorem{theorem}{Theorem}[section]

\newtheorem{cor}[theorem]{Corollary}

\theoremstyle{definition}

\theoremstyle{remark}

\usepackage{authblk} 
\usepackage{orcidlink} 
\usepackage{lineno}

\usetikzlibrary{arrows.meta, positioning}

\title{A mathematical model of HPAI transmission between dairy cattle and wild birds with environmental effects}

\author[1]{H.O.~Fatoyinbo\,\orcidlink{0000-0002-6036-2957}\thanks{Corresponding author: \href{mailto:hammed.fatoyinbo@aut.ac.nz}{hammed.fatoyinbo@aut.ac.nz}}}
\author[1]{P.~Tiwari\,\orcidlink{0000-0003-0206-6604}}
\author[2]{P.O.~Olanipekun\,\orcidlink{0000-0002-3310-1655}}
\author[3]{I.~Ghosh\,\orcidlink{0000-0001-8078-0577}}

\affil[1]{Department of Mathematical Sciences, Auckland University of Technology, Auckland 1010, New Zealand}
\affil[2]{Department of Mathematics, University of Auckland, Auckland 1010, New Zealand}
\affil[3]{School of Mathematics and Statistics, University College Dublin, Dublin, 4-D04 V1W8, Ireland}

\date{\today}

\begin{document}

\maketitle

\begin{abstract}

Highly pathogenic avian influenza (HPAI), particularly H5N1, poses an
increasing threat at the wildlife--livestock--environment interface,
with recent detections in dairy cattle motivating multi-host
transmission models. We develop and analyse a deterministic
compartmental model for HPAI transmission among cattle, wild birds,
and a shared contaminated environment. The model combines an SEIR
structure for cattle with an SIR framework for wild birds and includes
environmentally mediated transmission. We establish positivity and
boundedness, characterise the disease-free and endemic equilibria, and
derive the basic reproduction number $\mathcal R_0$ as the spectral
radius of a reduced next-generation matrix representing coupled
transmission pathways. We also establish threshold and global
stability properties of the equilibria. Numerical simulations
illustrate extinction, persistence, and parameter-dependent disease
dynamics. Global sensitivity analysis identifies cattle recovery,
cattle-to-cattle transmission, environmental viral decay, and
environmental transmission as the principal determinants of
$\mathcal R_0$ within the parameter ranges considered.
Response-surface and intervention analyses indicate that reducing
cattle-to-cattle transmission produces the greatest reduction in
transmission potential and peak cattle infection, while increasing
cattle recovery provides an additional effective pathway. Reducing
environmental transmission alone has a comparatively small effect
under the baseline parameterisation. Within the model assumptions and parameter ranges considered, the
findings provide qualitative insights into the relative effects of
transmission pathways and potential control measures.

\end{abstract}

\section{Introduction}

Avian influenza A viruses are commonly classified according to their pathogenicity into highly pathogenic (HPAI) and low pathogenic (LPAI) strains~\cite{LuSt15, SiRe25, cdcAvianInfluenza, HaMc25}. While LPAI infections are typically mild or subclinical, HPAI strains are associated with severe disease and high mortality in avian hosts. These viruses are further characterised by their surface glycoproteins, haemagglutinin (H) and neuraminidase (N), which define multiple viral subtypes. Among these, the H5N1 subtype has attracted considerable attention due to its broad host range, zoonotic potential, and capacity for rapid molecular evolution~\cite{WOAH_AvianInfluenza, stanislawek2024, Fatoyinbo2025}. Wild migratory waterfowl serve as natural reservoirs, facilitating long‑distance dissemination of the virus and transmission to domestic poultry such as chickens, turkeys, and geese~\cite{Nguyen2020H5Vietnam, stanislawek2024, Fatoyinbo2025}. Infected  wild birds shed virus through salivary, nasal, and fecal secretions, contaminating shared environments and enabling transmission via both direct contact and indirect exposure~\cite{BlTr21, GiJa24, SoHe25, cdcAvianInfluenza, cdcAvianInfluenza2}. Large-scale HPAI outbreaks have resulted in substantial losses to poultry production through mass culling, trade restrictions, and supply-chain disruptions, with serious implications for food security and rural livelihoods. In addition to their economic impact, H5Nx HPAI viruses are recognised zoonotic pathogens capable of infecting humans and other mammals, posing an ongoing public health concern. Consequently, HPAI outbreaks represent a multifaceted threat spanning animal health, food systems, and public health.

In recent years, substantial progress has been made in understanding avian influenza dynamics using genomic, molecular, and statistical approaches~\cite{KiJe24, DuIs22, MoAb24, LaBa23, Fereidouni23, vreman23, sangrat24, Esaki25}. Genomic surveillance has revealed frequent mutation and reassortment events within wild bird reservoirs, highlighting the adaptive potential of HPAI viruses and their propensity for cross‑species spillover. Statistical and epidemiological studies have further demonstrated pronounced spatial and temporal heterogeneity in outbreak occurrence. For example, Stanislawek \emph{et al.}~\cite{stanislawek2024} assessed the risk of H5 and H7 subtypes in wild birds in New Zealand through an integrated genomic and statistical framework, while Fatoyinbo \emph{et al.}~\cite{Fatoyinbo2025} developed a risk‑based statistical model to characterise HPAI H5 outbreaks across Asian subregions. While these approaches provide valuable empirical insight, they often do not resolve the underlying transmission mechanisms linking wildlife reservoirs, environmental persistence, and spillover into non‑avian hosts.

Beyond avian populations, HPAI viruses have increasingly been detected in a range of mammalian species. Infections with H5N1 have been documented in wild foxes~\cite{BoVr23, NiCh24} and other mammals, including domestic cats, dogs, and leopards~\cite{PlGa24}. Of particular interest is the recent documentation of HPAI H5N1 infections in dairy cattle~\cite{mpi1, SaBo25, RoPi24}, representing a novel host association. Genomic analyses indicate that these cattle infections are associated with specific H5N1 genotypes within clade 2.3.4.4b, which exhibit mammal-adaptive mutations and appear distinct from strains traditionally circulating in poultry and wild birds. Infected cattle generally exhibit mild or subclinical respiratory symptoms~\cite{CaFr24, cdcAvianInfluenza3}, with epidemiological investigations suggesting exposure through virus‑contaminated environments shared with wild birds. Experimental studies, such as those by Kalthoff \emph{et al.}~\cite{KaHo08}, have further demonstrated the susceptibility of calves to H5N1 under controlled conditions, supporting the biological plausibility of such spillover events.
These developments highlight important gaps in the theoretical understanding of avian influenza transmission across multiple host species. Mathematical modelling has long served as a powerful tool for elucidating the dynamics of infectious diseases such as dengue fever~\cite{Abidemi2022Wolbachia}, HIV~\cite{Deng2024HIVModel}, malaria~\cite{Rajnarayanan2025MalariaModel}, COVID‑19~\cite{AbidemiCovid}, and Lassa fever~\cite{Rejuetal24}. However, many existing avian influenza models focus primarily on poultry or wildlife reservoirs in isolation, often neglecting livestock hosts or treating environmental transmission in a simplified manner. Although previous modelling studies have examined human–avian transmission~\cite{IwTa07, WaFe08}, migratory processes~\cite{BoGo11}, delayed transmission dynamics~\cite{ShMo18}, and stochastic effects~\cite{PaBu13, HaMc25}, relatively few frameworks explicitly incorporate cattle as a host and examine the role of environmental contamination in mediating cross‑species spillover \cite{Rawson2025, CHANG2026}.

Motivated by these gaps, we develop a deterministic compartmental model to investigate the coupled transmission dynamics of HPAI among wild birds, cattle, and a shared environmental reservoir. The model integrates an SEIR structure for cattle with an SIR framework for wild birds and explicitly accounts for indirect transmission mediated by environmental viral persistence. Rather than focusing on outbreak reconstruction or parameter calibration, the model is designed to explore qualitative transmission mechanisms and threshold behaviour under homogeneous mixing assumptions.

The main contributions of this study are threefold. First, we formulate a multi‑host mathematical model that captures key interactions between wild birds, cattle, and environmental viral load. Second, we derive the basic reproduction number using the next‑generation matrix approach and analyse its dependence on epidemiological and environmental parameters through rigorous analytical and numerical techniques. Third, we establish the global stability properties of the disease‑free and endemic equilibria and use numerical simulations and global sensitivity analysis to examine how direct, cross‑species, and environmental pathways shape system dynamics. Together, these results provide a mathematically tractable framework for analysing HPAI transmission at the wildlife–livestock–environment interface and lay the groundwork for future extensions incorporating additional biological structure or control mechanisms.

\section{Model formulation} \label{sec2}

We propose a deterministic compartmental model to describe the transmission dynamics of highly pathogenic avian influenza (HPAI) at the cattle--wild bird--environment interface. The model is formulated at a population level under homogeneous mixing assumptions and incorporates indirect transmission through a shared environmental compartment $B(t)$ representing aggregate viral contamination from infected hosts. All parameters are assumed constant in time to facilitate analysis of qualitative dynamics and threshold behaviour.

\myEdit{
The homogeneous mixing assumption is adopted as a mathematically tractable, population-level simplification for examining the principal transmission pathways and their influence on threshold dynamics. Under this framework, the
transmission parameters are interpreted as population-averaged effective rates rather than as rates arising from uniformly distributed physical contacts among individuals. Specifically, $\beta_{bc}S_cI_b$ represents effective
cross-species transmission from infected wild birds to susceptible cattle. Such transmission may occur where wild-bird and cattle populations overlap around grazing areas, water sources, feed-storage sites, barns, or other shared farm environments. This mechanism is distinguished from environmentally mediated transmission, represented by $\beta_{\mathrm{env}}S_cB$, which captures exposure to accumulated viral contamination in the shared environment.

In practice, contact patterns among cattle are shaped by herd structure and shared management facilities, while interactions between wild birds and cattle are influenced by spatial clustering, resource use, seasonal variation, and
migratory movements. Consequently, the model is designed to capture the principal transmission processes at the population level in a mathematically tractable manner rather than to reproduce farm-specific or spatially explicit contact networks.
}

\begin{figure}[ht]
\centering
\includegraphics[width=0.75\linewidth]{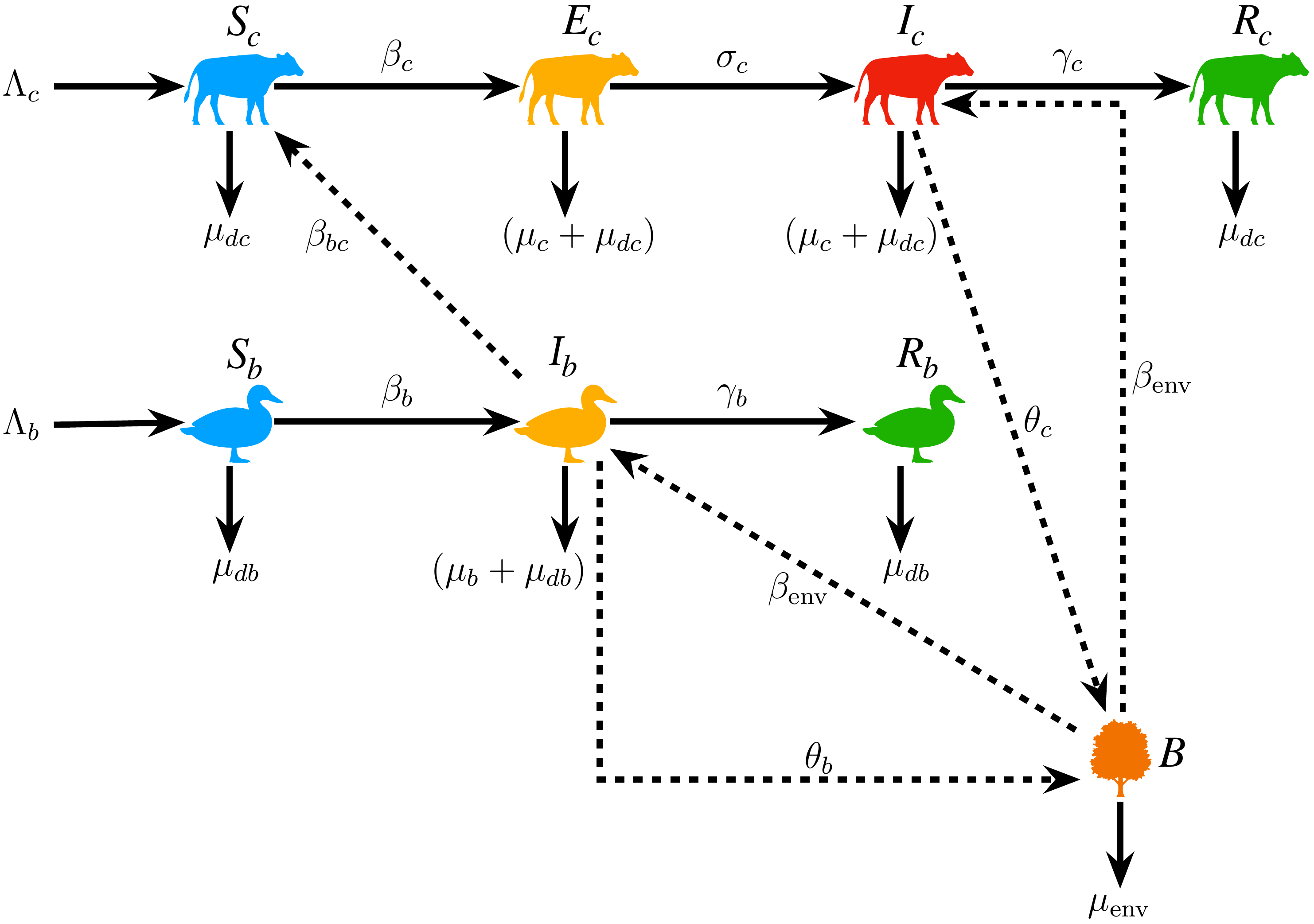}
\caption{Schematic diagram of HPAI transmission among cattle, wild birds, and the environment}
\label{fig:compartment}
\end{figure}

Cattle are divided into susceptible ($S_c$), exposed ($E_c$), infected ($I_c$), and recovered ($R_c$) compartments, while wild birds are classified as susceptible ($S_b$), infected ($I_b$), or recovered ($R_b$). The transmission pathways connecting host populations and the environment are illustrated in Figure~\ref{fig:compartment}.

Cattle are recruited into the susceptible class at rate $\Lambda_c$.
Susceptible cattle become exposed through effective contact with infected
cattle, infected wild birds, or the contaminated environment. Exposed cattle
progress to the infectious class at rate $\sigma_c$. Infected cattle recover at rate $\gamma_c$ or are removed through natural
mortality at rate $\mu_{dc}$ and disease-induced mortality at rate $\mu_c$.

\myEdit{
Here, the distinction between $E_c$ and $I_c$ is based on infectiousness rather
than the presence of clinical signs. The class $E_c$ represents infected cattle
that have not yet begun shedding infectious virus, whereas $I_c$ includes all
cattle capable of transmission or environmental shedding, including animals
that may be preclinical or asymptomatic. Thus, $\sigma_c$ represents
progression from the non-infectious latent stage to the infectious stage and
does not necessarily correspond to the onset of observable clinical disease.
}

Wild birds are recruited into the susceptible class at rate $\Lambda_b$. \myEdit{This recruitment term includes susceptible birds only; immigration of
infected migratory birds and external introduction of environmental contamination are not represented. Consequently, the model describes local transmission dynamics in the absence of repeated viral importation, and the
disease-free equilibrium corresponds to local elimination conditional on no subsequent reintroduction.}

Susceptible wild birds acquire infection through direct transmission from infected wild birds or through exposure to the contaminated environment. Infected wild birds recover at rate $\gamma_b$ or are removed through natural mortality at rate $\mu_{db}$ and disease-induced mortality at rate $\mu_b$.

The environmental compartment $B(t)$ represents a lumped, scaled measure of
aggregate infectious viral contamination in the shared environment that is
potentially accessible to cattle and wild birds, rather than a directly
measured concentration in a specific environmental medium. It increases
through viral shedding from infected cattle and wild birds at rates
$\theta_c$ and $\theta_b$, respectively, and decreases through viral decay or
removal at rate $\mu_{\mathrm{env}}$.

\myEdit{
Accordingly, $\theta_c$, $\theta_b$, and $\beta_{\mathrm{env}}$ are effective
parameters defined relative to the chosen scale of $B(t)$, while
$\mu_{\mathrm{env}}$ represents an effective average decay or removal rate
rather than a universal physical constant.

More detailed avian-influenza models distinguish environmental viral loads by
host and location and incorporate seasonally varying decay rates
\cite{HaMc25}. In the present study, these processes are aggregated into a
single environmental compartment with a constant effective decay rate to
preserve analytical tractability. This approximation does not imply that viral
persistence is constant under field conditions, where it may vary with
temperature, humidity, ultraviolet radiation, salinity, pH, viral strain, and
environmental substrate \cite{Martin2018,HaMc25}.
}

For compactness, we define the forces of infection for cattle and wild birds as
\begin{equation}
\label{eqn:lambda_values}
\begin{aligned}
\lambda_c=&\beta_c I_c+\beta_{bc}I_b+\beta_{\mathrm{env}}B,\\
\lambda_b&=\beta_b I_b+\beta_{\mathrm{env}}B.
\end{aligned}
\end{equation}
Thus, $\lambda_c$ represents the total infection pressure on susceptible cattle due to infected cattle, infected wild birds, and the contaminated environment, while $\lambda_b$ represents the infection pressure on susceptible wild birds due to infected wild birds and the contaminated environment. Equivalently,
\[
\lambda_c S_c
=
\beta_c S_c I_c+\beta_{bc}S_c I_b+\beta_{\mathrm{env}}S_cB,
\qquad
\lambda_b S_b
=
\beta_b S_b I_b+\beta_{\mathrm{env}}S_bB.
\]

The full SEIR--SIR--environment model is given by
\begin{equation}
\label{eqn:fullmodel}
\begin{aligned}
\frac{dS_c}{dt} 
&= \Lambda_c-\lambda_c S_c-\mu_{dc}S_c, \\
\frac{dE_c}{dt} 
&= \lambda_c S_c-(\sigma_c+\mu_c+\mu_{dc})E_c, \\
\frac{dI_c}{dt} 
&= \sigma_cE_c-(\gamma_c+\mu_c+\mu_{dc})I_c, \\
\frac{dR_c}{dt} 
&= \gamma_cI_c-\mu_{dc}R_c, \\
\frac{dS_b}{dt} 
&= \Lambda_b-\lambda_b S_b-\mu_{db}S_b, \\
\frac{dI_b}{dt} 
&= \lambda_b S_b-(\gamma_b+\mu_b+\mu_{db})I_b, \\
\frac{dR_b}{dt} 
&= \gamma_bI_b-\mu_{db}R_b, \\
\frac{dB}{dt} 
&= \theta_cI_c+\theta_bI_b-\mu_{\mathrm{env}}B .
\end{aligned}
\end{equation}

The recovered compartments $R_c$ and $R_b$ do not feed back into the transmission dynamics. \myEdit{In the present formulation, recovered cattle and wild birds are assumed to remain immune over the time period considered and therefore do not return to the susceptible classes. This is a simplifying assumption rather than a claim of biologically established lifelong immunity, since the duration of post-infection immunity to HPAI H5N1 in cattle and wild birds remains uncertain.} Their equations are linearly driven by $I_c$ and $I_b$, respectively. Hence, the subsystem for $(S_c,E_c,I_c,S_b,I_b,B)$ is closed. Consequently, the theoretical analysis can be carried out on the reduced system without loss of generality, while $R_c$ and $R_b$ can be recovered from the corresponding linear equations in~\eqref{eqn:fullmodel}.

The reduced system used for the analytical results is therefore
\begin{equation}
\label{eqn:reduced}
\begin{aligned}
\frac{dS_c}{dt} 
&= \Lambda_c-\lambda_c S_c-\mu_{dc}S_c, \\
\frac{dE_c}{dt} 
&= \lambda_c S_c-(\sigma_c+\mu_c+\mu_{dc})E_c, \\
\frac{dI_c}{dt} 
&= \sigma_cE_c-(\gamma_c+\mu_c+\mu_{dc})I_c, \\
\frac{dS_b}{dt} 
&= \Lambda_b-\lambda_b S_b-\mu_{db}S_b, \\
\frac{dI_b}{dt} 
&= \lambda_b S_b-(\gamma_b+\mu_b+\mu_{db})I_b, \\
\frac{dB}{dt} 
&= \theta_cI_c+\theta_bI_b-\mu_{\mathrm{env}}B .
\end{aligned}
\end{equation}

A complete description of the model parameters is provided in Table~\ref{tab:paramdescr}. From this point onward, all analytical and numerical investigations are conducted on the reduced system~\eqref{eqn:reduced}.

\begin{table}[h!]
\centering
\caption{Model parameters and their descriptions}
\label{tab:paramdescr}
\begin{tabular}{|c|p{10cm}|}
\hline
\textbf{Parameter} & \textbf{Description} \\
\hline\hline
$\Lambda_b$ & Recruitment rate of wild birds. \\
$\Lambda_c$ & Recruitment rate of cattle. \\
$\beta_b$ & Effective transmission rate among wild birds. \\
$\beta_c$ & Effective transmission rate among cattle. \\
$\beta_{bc}$ &Effective cross-species transmission rate from infected wild birds to susceptible cattle. \\
$\beta_{\text{env}}$ & Effective transmission coefficient associated with exposure to contaminated environment. \\
$\sigma_c$ & Progression rate from exposed to infected cattle. \\
$\mu_c$ & Disease--induced mortality rate in cattle. \\
$\mu_b$ & Disease--induced mortality rate in birds. \\
$\mu_{dc}$ & Natural mortality rate in cattle. \\
$\mu_{db}$ & Natural mortality rate in birds. \\
$\mu_{\text{env}}$ & Effective average decay or removal rate of virus in the environment. \\
$\gamma_b$ & Recovery rate of infected birds. \\
$\gamma_c$ & Recovery rate of infected cattle. \\
$\theta_b$ & Effective environmental contamination rate by infected birds. \\
$\theta_c$ & Effective environmental contamination rate by infected cattle. \\
\hline
\end{tabular}
\end{table}

\section{Qualitative properties of the model}\label{sec3}

In this section, we investigate the fundamental qualitative properties of system~\eqref{eqn:reduced}. Specifically, we establish positivity and boundedness of solutions, determine the disease--free and endemic equilibria, derive the basic reproduction number $\mathcal{R}_0$ using the next--generation matrix approach, and analyse the stability of equilibria to characterise disease persistence. In addition, we assess the sensitivity of $\mathcal{R}_0$ to model parameters using Latin Hypercube Sampling (LHS) and Partial Rank Correlation Coefficients (PRCC)~\cite{ABIDEMI2022}.

\subsection{Positivity and boundedness}

Since the model describes interacting populations of cattle, wild birds, and an environmental viral reservoir, it is necessary to verify that the solutions remain epidemiologically meaningful. In particular, we require that all state variables remain non‑negative and bounded for all $t \ge 0$. Let
\[
N_c(t)=S_c(t)+E_c(t)+I_c(t)+R_c(t),
\qquad
N_b(t)=S_b(t)+I_b(t)+R_b(t)
\]
denote the total cattle and wild-bird populations, respectively. Define
\[
\mathcal E_c:=\left\{(S_c,E_c,I_c,R_c)\in\mathbb R_+^4:
N_c\leq \frac{\Lambda_c}{\mu_{dc}}\right\},
\quad\quad
\mathcal E_b:=\left\{(S_b,I_b,R_b)\in\mathbb R_+^3:
N_b\leq \frac{\Lambda_b}{\mu_{db}}\right\},
\]
and
\[
\mathcal B:=\left\{B\in\mathbb R_+:
B\leq \frac{1}{\mu_{\mathrm{env}}}
\left(
\theta_c\frac{\Lambda_c}{\mu_{dc}}+
\theta_b\frac{\Lambda_b}{\mu_{db}}
\right)\right\}.
\]
The biologically feasible region for the full system is therefore
\[
\mathcal E:=\mathcal E_c\times\mathcal E_b\times\mathcal B.
\]
Its projection onto $(S_c,E_c,I_c,S_b,I_b,B)$ gives the corresponding feasible region for the reduced system ~\eqref{eqn:reduced}.

\begin{theorem}\label{thm1}
Suppose that the initial conditions satisfy
\[
S_c(0),E_c(0),I_c(0),R_c(0),S_b(0),I_b(0),R_b(0),B(0)\geq 0.
\]
Then every solution of the model that starts in the non-negative orthant remains non-negative for all $t\geq 0$.
\end{theorem}

\begin{theorem}\label{thm:1}
The cattle population, wild-bird population, and environmental viral load are bounded from above for all non-negative initial data.
\end{theorem}

The proofs of theorems~\ref{thm1} and~\ref{thm:1} are provided in the Appendix.

\begin{cor}
Suppose that
\[
N_c(0)\leq \frac{\Lambda_c}{\mu_{dc}},
\qquad
N_b(0)\leq \frac{\Lambda_b}{\mu_{db}},
\]
and
\[
B(0)\leq
\frac{1}{\mu_{\mathrm{env}}}
\left(
\theta_c\frac{\Lambda_c}{\mu_{dc}}+
\theta_b\frac{\Lambda_b}{\mu_{db}}
\right).
\]
Then the region $\mathcal E$ is positively invariant.
\end{cor}

\begin{proof}
From the upper comparison inequalities in the proof of Theorem \ref{thm:1},
\[
N_c(t)\leq \frac{\Lambda_c}{\mu_{dc}},
\qquad
N_b(t)\leq \frac{\Lambda_b}{\mu_{db}}
\qquad \text{for all }t\geq 0.
\]
Consequently,
\[
\frac{dB}{dt}
\leq
\theta_c\frac{\Lambda_c}{\mu_{dc}}
+\theta_b\frac{\Lambda_b}{\mu_{db}}
-\mu_{\mathrm{env}}B.
\]
Hence
\[
B(t)\leq
\frac{1}{\mu_{\mathrm{env}}}
\left(
\theta_c\frac{\Lambda_c}{\mu_{dc}}
+\theta_b\frac{\Lambda_b}{\mu_{db}}
\right)
+
\left[
B(0)-
\frac{1}{\mu_{\mathrm{env}}}
\left(
\theta_c\frac{\Lambda_c}{\mu_{dc}}
+\theta_b\frac{\Lambda_b}{\mu_{db}}
\right)
\right]e^{-\mu_{\mathrm{env}}t}.
\]
Therefore $B(t)$ remains in $\mathcal B$, and together with Theorem \ref{thm1} this shows that $\mathcal E$ is positively invariant.
\end{proof}

\subsection{Equilibrium analysis}
An equilibrium point of a dynamical system is a point on the state space at which the system reaches an equilibrium, that is, the dynamics do not change further. The first step towards interpreting the intricate properties of a dynamical system is to analyse its equilibrium points. Epidemiological models often admit a disease-free equilibrium and, under suitable parameter conditions, one or more endemic equilibria. We first determine the disease-free equilibrium and then characterise the endemic equilibrium structure of system~\eqref{eqn:reduced}. 

First, we compute the disease-free equilibrium (DFE), which occurs when there is no infection in the population. This implies that all compartments related to infection $E_{c}, I_{c}, I_{b}$ and $B$ are zero. Therefore the DFE, denoted $\psi_{\text{DFE}}$, for the model \eqref{eqn:reduced} is 
\begin{equation}
\label{eq:DFE}
\psi_{\text{DFE}}(S_{c}^{*},E_{c}^{*}, I_{c}^{*},S_{b}^{*}, I_{b}^{*},B^{*})=\left(\frac{\Lambda_c}{\mu_{dc}}, 0, 0,\frac{\Lambda_b}{\mu_{db}}, 0, 0\right).
\end{equation}
We will use $\psi_{\rm DFE}$ to compute the {\em basic reproduction number} $\mathcal{R}_0$. \myEdit{It quantifies the
asymptotic number of secondary infections generated through the
coupled cattle, wild-bird, and environmental transmission pathways
when infection is introduced into an otherwise susceptible system.
The threshold $\mathcal R_0=1$ distinguishes parameter regimes in
which infection can invade from those in which it cannot persist in
the absence of further external introduction}. In order to compute $\mathcal{R}_0$, we look into the system of ODEs that represent the infected classes 
\begin{equation}
\begin{aligned}
\label{eqn:infected}
\frac{dE_c}{dt} &= \beta_c S_c I_c + \beta_{bc} S_c I_b + \beta_{\text{env}} S_c B - \sigma_c E_c-\mu_c E_c - \mu_{dc} E_{c}, \\
\frac{dI_c}{dt} &= \sigma_c E_c - \gamma_c I_c-\mu_c I_c -\mu_{dc} I_{c}, \\
\frac{dI_b}{dt} &= \beta_b S_b I_b  + \beta_{\text{env}} S_b B - \gamma_b I_b-\mu_b I_b - \mu_{db} I_{b}, \\
\frac{dB}{dt} &= \theta_c I_c + \theta_b I_b - \mu_{\text{env}} B.
\end{aligned}
\end{equation}
The infection matrix $\mathcal{F}$ describes the flow of new infections. It corresponds to the terms that generate new infections in the equations for $E_{c}, I_{c}, I_{b}$ and $B$. At $\psi_{\rm DFE}$, we have
\begin{equation}
\mathcal{F}=    \begin{pmatrix}
    0 &\beta_c \frac{\Lambda_c}{\mu_{dc}} & \beta_{bc} \frac{\Lambda_c}{\mu_{dc}} & \beta_{\text{env}} \frac{\Lambda_c}{\mu_{dc}}\\
    0 & 0 &0 &0\\
     0 & 0 & \beta_b \frac{\Lambda_b}{\mu_{db}}& \beta_{\text{env}} \frac{\Lambda_b}{\mu_{db}}\\
      0 & 0 & 0&0\\
    \end{pmatrix}
\end{equation}
The transition matrix $\mathcal{V}$ describes the flow of individuals out of infected compartments due to recovery or other processes. The transition matrix is derived from the negative terms in the equations. At $\psi_{\rm DFE}$ we have
\begin{equation*}
\mathcal{V}=    \begin{pmatrix}
    \sigma_c+\mu_c+\mu_{dc} & 0& 0 &0 \\
   -\sigma_c  & \gamma_c+\mu_c+ \mu_{dc} &0 &0\\
     0 & 0 & \gamma_b+\mu_b+\mu_{db}&0 \\
      0 & -\theta_c & -\theta_b&\mu_{\text{env}}\\
    \end{pmatrix}
\end{equation*}
To obtain the basic reproduction number $\mathcal{R}_{0}$, we use the next generation matrix (NGM) approach. We denote the NGM as $\xi_{\text{NGM}}=\mathcal{FV}^{-1}$. We have
\begin{equation*}
\mathcal{V}^{-1} = \begin{pmatrix}
\frac{1}{\sigma_c + \mu_c+\mu_{dc}} & 0 & 0 & 0 \\
\frac{\sigma_c}{(\sigma_c + \mu_c+\mu_{dc})(\gamma_c + \mu_c+\mu_{dc})} & \frac{1}{\gamma_c + \mu_c+\mu_{dc}} & 0 & 0 \\
0 & 0 & \frac{1}{\gamma_b + \mu_b+\mu_{db}} & 0 \\
\frac{\sigma_c \theta_c}{(\sigma_c + \mu_c+\mu_{dc})(\gamma_c + \mu_c+\mu_{dc})\mu_{\text{env}}} & \frac{\theta_c}{(\gamma_c + \mu_c+\mu_{dc})\mu_{\text{env}}} & \frac{\theta_b}{(\gamma_b + \mu_b+\mu_{db})\mu_{\text{env}}} & \frac{1}{\mu_{\text{env}}}
\end{pmatrix}
\end{equation*}
Then the NGM is given by
\begin{equation*}
 \xi_{\text{NGM}}=
\begin{pmatrix}
\dfrac{\Lambda_c \sigma_c (\beta_c \mu_{\text{env}} + \beta_{\text{env}} \theta_c)}{\mu_{dc} \mu_{\text{env}} D_c} & 
\dfrac{\Lambda_c (\beta_c \mu_{\text{env}} + \beta_{\text{env}} \theta_c)}{\mu_{dc} \mu_{\text{env}} (\gamma_c + \mu_c + \mu_{dc})} & 
\dfrac{\Lambda_c (\beta_{bc} \mu_{\text{env}} + \beta_{\text{env}} \theta_b)}{\mu_{dc} \mu_{\text{env}} (\gamma_b + \mu_b + \mu_{db})} & 
\dfrac{\Lambda_c \beta_{\text{env}}}{\mu_{dc} \mu_{\text{env}}}
\\[10pt]
0 & 0 & 0 & 0
\\[10pt]
\dfrac{\Lambda_b \beta_{\text{env}} \sigma_c \theta_c}{\mu_{db} \mu_{\text{env}} D_c} & 
\dfrac{\Lambda_b \beta_{\text{env}} \theta_c}{\mu_{db} \mu_{\text{env}} (\gamma_c + \mu_c + \mu_{dc})} & 
\dfrac{\Lambda_b \beta_b \mu_{\text{env}} + \Lambda_b \beta_{\text{env}} \theta_b}{\mu_{db} \mu_{\text{env}} (\gamma_b + \mu_b + \mu_{db})} & 
\dfrac{\Lambda_b \beta_{\text{env}}}{\mu_{db} \mu_{\text{env}}}
\\[10pt]
0 & 0 & 0 & 0
\end{pmatrix},
\end{equation*}

\myEdit{
where
\[
D_c
=
(\sigma_c+\mu_c+\mu_{dc})
(\gamma_c+\mu_c+\mu_{dc}).
\]

Since the second and fourth rows of
$\xi_{\mathrm{NGM}}=\mathcal F\mathcal V^{-1}$ are zero, the
next-generation matrix has two zero eigenvalues. Its two potentially
nonzero eigenvalues are the eigenvalues of the reduced matrix
\[
K
=
\begin{pmatrix}
K_{cc} & K_{cb}\\
K_{bc} & K_{bb}
\end{pmatrix},
\]
where
\begin{align*}
K_{cc}
&=
\frac{
\Lambda_c\sigma_c
\left(
\beta_c\mu_{\mathrm{env}}
+\beta_{\mathrm{env}}\theta_c
\right)
}{
\mu_{dc}\mu_{\mathrm{env}}D_c
},\\
K_{cb}
&=
\frac{
\Lambda_c
\left(
\beta_{bc}\mu_{\mathrm{env}}
+\beta_{\mathrm{env}}\theta_b
\right)
}{
\mu_{dc}\mu_{\mathrm{env}}
(\gamma_b+\mu_b+\mu_{db})
},\\
K_{bc}
&=
\frac{
\Lambda_b\beta_{\mathrm{env}}
\sigma_c\theta_c
}{
\mu_{db}\mu_{\mathrm{env}}D_c
},\\
K_{bb}
&=
\frac{
\Lambda_b
\left(
\beta_b\mu_{\mathrm{env}}
+\beta_{\mathrm{env}}\theta_b
\right)
}{
\mu_{db}\mu_{\mathrm{env}}
(\gamma_b+\mu_b+\mu_{db})
}.
\end{align*}

Therefore, the basic reproduction number is the spectral radius of
the next-generation matrix:
\begin{equation}
\mathcal R_0
=
\rho\left(\mathcal F\mathcal V^{-1}\right)
=
\rho(K).
\label{eq:R0_spectral}
\end{equation}
The eigenvalues of $K$ are
\[
\lambda_{\pm}
=
\frac{
K_{cc}+K_{bb}
\pm
\sqrt{
(K_{cc}-K_{bb})^2+4K_{cb}K_{bc}
}
}{2}.
\]
Since $K$ is a nonnegative matrix, its spectral radius is the larger
of these two eigenvalues. Hence,
\begin{equation}
\mathcal R_0
=
\frac{
K_{cc}+K_{bb}
+
\sqrt{
(K_{cc}-K_{bb})^2+4K_{cb}K_{bc}
}
}{2}.
\label{eq:R0}
\end{equation}

This formulation reflects the coupled transmission pathways among
cattle, wild birds, and the shared environment. The quantity $K_{cc}$
represents the expected number of newly infected cattle generated by
an infected-cattle lineage through direct cattle-to-cattle
transmission and cattle-origin environmental contamination. Similarly,
$K_{bb}$ represents the expected number of newly infected wild birds
generated by an infected-bird lineage through direct bird-to-bird
transmission and bird-origin environmental contamination.

The off-diagonal quantity $K_{cb}$ represents the expected number of
new cattle infections generated by an infected-bird lineage. It
includes direct wild-bird-to-cattle transmission through
$\beta_{bc}$ and indirect transmission arising from environmental
contamination by infected wild birds. In contrast, $K_{bc}$ represents the expected number of new
wild-bird infections generated by an infected-cattle lineage through
the environmental pathway. The model does not include a direct cattle-to-bird transmission term.

}

\myEdit{

\noindent
Next, we compute the endemic equilibrium of model \eqref{eqn:reduced}, which is represented as
\[
\psi_{\mathrm{EE}}
=
\left(
S_c^{**},E_c^{**},I_c^{**},
S_b^{**},I_b^{**},B^{**}
\right).
\]
At the endemic equilibrium, the infected classes are non-zero. Since the
equilibrium equations are strongly coupled through the environmental reservoir,
a closed-form expression is not generally available. We therefore reduce the
equilibrium conditions to a nonlinear system that can be solved numerically for
any specified parameter set.
\noindent
Setting the right-hand side of model \eqref{eqn:reduced} equal to zero gives
\begin{align*}
\text{(i)}\qquad
B^{**}
&=
\frac{\theta_c I_c^{**}+\theta_b I_b^{**}}{\mu_{\mathrm{env}}},
\\
\text{(ii)}\qquad
I_b^{**}
&=
\frac{\beta_{\mathrm{env}}S_b^{**}B^{**}}
{\mu_{db}+\mu_b+\gamma_b-\beta_bS_b^{**}},
\\
\text{(iii)}\qquad
S_b^{**}
&=
\frac{\Lambda_b}
{\beta_b I_b^{**}+\beta_{\mathrm{env}}B^{**}+\mu_{db}},
\\
\text{(iv)}\qquad
I_c^{**}
&=
\frac{\sigma_c E_c^{**}}
{\gamma_c+\mu_c+\mu_{dc}},
\\
\text{(v)}\qquad
E_c^{**}
&=
\frac{
\beta_c S_c^{**}I_c^{**}
+\beta_{bc}S_c^{**}I_b^{**}
+\beta_{\mathrm{env}}S_c^{**}B^{**}
}
{\sigma_c+\mu_c+\mu_{dc}},
\\
\text{(vi)}\qquad
S_c^{**}
&=
\frac{\Lambda_c}
{\beta_c I_c^{**}
+\beta_{bc} I_b^{**}
+\beta_{\mathrm{env}}B^{**}
+\mu_{dc}}.
\end{align*}

\noindent
Now utilising the last three equations (iv)--(vi), we obtain
\begin{equation}
I_c^{**}
=
\frac{\sigma_c\Lambda_c}
{(\gamma_c+\mu_c+\mu_{dc})(\sigma_c+\mu_c+\mu_{dc})}
\left[
\frac{
\beta_c I_c^{**}+f_1(I_c^{**},B^{**})
}{
\beta_c I_c^{**}+f_1(I_c^{**},B^{**})+\mu_{dc}
}
\right],
\label{eq:3.9}
\end{equation}
where
\[
f_1(I_c^{**},B^{**})
=
\beta_{bc}I_b^{**}
+\beta_{\mathrm{env}}B^{**}.
\]

\noindent
To express $I_b^{**}$ in terms of $I_c^{**}$ and $B^{**}$, equations (ii) and
(iii) give
\[
I_b^{**}
=
\frac{
\beta_{\mathrm{env}}\Lambda_b B^{**}
}{
(\mu_{db}+\mu_b+\gamma_b)
\left(
\beta_b I_b^{**}
+\beta_{\mathrm{env}}B^{**}
+\mu_{db}
\right)
-\beta_b\Lambda_b
}.
\]
Also, from equation (i),
\[
I_b^{**}
=
\frac{\mu_{\mathrm{env}}B^{**}-\theta_c I_c^{**}}{\theta_b}.
\]
Substituting this expression into the denominator above yields
\[
I_b^{**}
=
\frac{
\beta_{\mathrm{env}}\Lambda_b\theta_b B^{**}
}{
f_2(I_c^{**},B^{**})
},
\]
where
\begin{align}
f_2(I_c^{**},B^{**})
={}&
(\mu_{db}+\mu_b+\gamma_b)
\Big(
\beta_b\mu_{\mathrm{env}}B^{**}
-\beta_b\theta_c I_c^{**}
+\theta_b\beta_{\mathrm{env}}B^{**}
+\theta_b\mu_{db}
\Big)
\nonumber\\
&\quad
-\beta_b\Lambda_b\theta_b.
\label{3.10a}
\end{align}
Hence
\begin{align}
f_1(I_c^{**},B^{**})
&=
\beta_{bc}I_b^{**}
+\beta_{\mathrm{env}}B^{**}
\nonumber\\
&=
\left[
\frac{
\beta_{bc}\Lambda_b\theta_b
}{
f_2(I_c^{**},B^{**})
}
+1
\right]
\beta_{\mathrm{env}}B^{**}.
\label{3.10}
\end{align}
Let
\[
A
=
\frac{\sigma_c\Lambda_c}
{(\gamma_c+\mu_c+\mu_{dc})(\sigma_c+\mu_c+\mu_{dc})},
\]
and define
\[
f_3(I_c^{**},B^{**})
=
\beta_c I_c^{**}f_2(I_c^{**},B^{**})
+
\left\{
\beta_{bc}\Lambda_b\theta_b
+
f_2(I_c^{**},B^{**})
\right\}
\beta_{\mathrm{env}}B^{**}.
\]
Then
\[
\beta_c I_c^{**}+f_1(I_c^{**},B^{**})
=
\frac{
f_3(I_c^{**},B^{**})
}{
f_2(I_c^{**},B^{**})
}.
\]
Therefore, equation \eqref{eq:3.9} becomes
\begin{equation}
I_c^{**}
\left[
f_3(I_c^{**},B^{**})
+\mu_{dc}f_2(I_c^{**},B^{**})
\right]
-
A f_3(I_c^{**},B^{**})
=
0.
\label{eq:3.11}
\end{equation}
Equation \eqref{eq:3.11} alone does not close the system because both $I_c^{**}$ and
$B^{**}$ are unknown. A second relation follows by equating the two expressions
for $I_b^{**}$ obtained above:
\[
\frac{\mu_{\mathrm{env}}B^{**}-\theta_c I_c^{**}}{\theta_b}
=
\frac{
\beta_{\mathrm{env}}\Lambda_b\theta_b B^{**}
}{
f_2(I_c^{**},B^{**})
}.
\]
Thus,
\begin{equation}
\left(
\mu_{\mathrm{env}}B^{**}-\theta_c I_c^{**}
\right)
f_2(I_c^{**},B^{**})
-
\beta_{\mathrm{env}}\Lambda_b\theta_b^2 B^{**}
=
0.
\label{eq:3.12}
\end{equation}
Hence, the endemic equilibrium is determined by solving the coupled nonlinear
equations \eqref{eq:3.11}--\eqref{eq:3.12} for $I_c^{**}>0$ and $B^{**}>0$. Once
$I_c^{**}$ and $B^{**}$ are obtained, $I_b^{**}$, $S_b^{**}$,
$E_c^{**}$, and $S_c^{**}$ follow directly from equations (i)--(vi).
}

\subsection{Global stability analysis of equilibria}

Here we establish the global stability properties of the
disease-free and endemic equilibria. The disease-free equilibrium is
analysed using a linear Lyapunov function and the theory of positive
systems, whereas the endemic equilibrium is analysed using a
Volterra-type Lyapunov function. The detailed calculation is given in Appendix~\ref{sec:GlobStab}.

\myEdit{\begin{theorem}[Threshold and stability properties]
\label{thm:threshold_stability}
Consider system~\eqref{eqn:reduced} in the biologically feasible
region $\mathcal E$, and let
\[
\mathcal R_0=\rho(\mathcal F\mathcal V^{-1}).
\]
Then the following statements hold:
\begin{enumerate}
\item If $\mathcal R_0<1$, the disease-free equilibrium
$\psi_{\text{DFE}}$ is globally asymptotically stable in $\mathcal E$.

\item If $\mathcal R_0>1$, the disease-free equilibrium
$\psi_{\text{DFE}}$ is unstable.

\item Suppose that the model admits a unique strictly positive
endemic equilibrium $\psi_{\text{EE}}$ and that all transmission and
shedding parameters are strictly positive. Then
$\psi_{\text{EE}}$ is globally asymptotically stable in the interior
of $\mathcal E$.
\end{enumerate}
In the absence of external viral introduction, these results describe
local elimination or persistence within the modelled system.
\end{theorem}}

\subsection{Global sensitivity analysis}
\label{sec:GSA}

To assess the influence of model parameters on transmission potential,
we performed a global sensitivity analysis of the basic reproduction
number $\mathcal R_0$ using Latin Hypercube Sampling (LHS) in
combination with Partial Rank Correlation Coefficients (PRCC). This
approach permits systematic exploration of parameter uncertainty and
quantifies monotonic associations between the input parameters and the
model output while controlling for the effects of the remaining
sampled parameters.

A total of $N=5000$ parameter sets were generated using LHS.
The parameters were sampled from uniform distributions over
biologically and epidemiologically plausible ranges. The use of
uniform distributions reflects limited prior information and permits
broad exploration of the specified parameter ranges, consistent with
the qualitative and exploratory purpose of the analysis.

The sensitivity analysis focused on parameters directly influencing
transmission, progression, recovery, environmental contamination, and
viral persistence, namely
\[
\beta_c,\quad
\beta_b,\quad
\beta_{bc},\quad
\beta_{\mathrm{env}},\quad
\sigma_c,\quad
\gamma_c,\quad
\gamma_b,\quad
\theta_c,\quad
\theta_b,\quad
\mu_{\mathrm{env}}.
\]
Demographic parameters, including the recruitment and natural
mortality or effective demographic removal rates, were held fixed at
their baseline values. For each sampled parameter set, the
corresponding value of $\mathcal R_0$ was calculated using the
spectral-radius expression in Equation~\eqref{eq:R0}.
\myEdit{
Let
\[
X_j
=
\left(
X_{1j},X_{2j},\ldots,X_{Nj}
\right)^{\mathsf T}
\]
denote the sampled values of parameter $j$, and let
\[
Y
=
\left(
\mathcal R_0^{(1)},\mathcal R_0^{(2)},\ldots,
\mathcal R_0^{(N)}
\right)^{\mathsf T}
\]
denote the corresponding values of the basic reproduction number.
The parameter values and model outputs were first replaced by their
ranks:
\[
U_j=\operatorname{rank}(X_j),
\qquad
V=\operatorname{rank}(Y).
\]

For parameter $j$, let $U_{-j}$ denote the matrix containing the
ranked values of all sampled parameters except parameter $j$. To
remove the effects of the remaining parameters, the ranked parameter
$U_j$ and ranked model output $V$ were separately regressed on
$U_{-j}$:
\begin{align}
U_j
&=
a_j+U_{-j}\boldsymbol{\alpha}_j+\varepsilon_j,
\label{eq:prcc_parameter_regression}\\
V
&=
b_j+U_{-j}\boldsymbol{\beta}_j+\eta_j,
\label{eq:prcc_output_regression}
\end{align}
where $\varepsilon_j$ and $\eta_j$ are the corresponding residual
vectors.

The PRCC for parameter $j$ was then calculated as the Pearson
correlation coefficient between these residual vectors:
\begin{equation}
\operatorname{PRCC}_j
=
\operatorname{Corr}(\varepsilon_j,\eta_j)
=
\frac{
\displaystyle
\sum_{i=1}^{N}
\left(\varepsilon_{ij}-\bar{\varepsilon}_j\right)
\left(\eta_{ij}-\bar{\eta}_j\right)
}{
\displaystyle
\sqrt{
\sum_{i=1}^{N}
\left(\varepsilon_{ij}-\bar{\varepsilon}_j\right)^2
\sum_{i=1}^{N}
\left(\eta_{ij}-\bar{\eta}_j\right)^2
}
}.
\label{eq:prcc}
\end{equation}

Thus, $\operatorname{PRCC}_j$ measures the monotonic association
between parameter $j$ and $\mathcal R_0$ after controlling for the
ranked values of the remaining sampled parameters. PRCC values lie in
the interval $[-1,1]$.} A positive PRCC indicates that
$\mathcal R_0$ tends to increase as the parameter increases, after
adjusting for the other parameters, whereas a negative PRCC indicates
that $\mathcal R_0$ tends to decrease. Values with larger absolute
magnitudes indicate stronger adjusted monotonic associations.

\section{Numerical simulation and results}\label{sec4}

In this section, we explore the qualitative dynamical behaviour of system~\eqref{eqn:reduced} through numerical simulations. All simulations were performed in MATLAB using adaptive ODE solvers with prescribed relative and absolute tolerances. The numerical experiments are intended to illustrate generic transmission patterns, threshold behaviour, and the influence of parameter uncertainty, rather than to reproduce or forecast a specific outbreak.

Unless otherwise stated, simulations were initiated from the reference initial conditions
\[
S_c(0)=1499,\quad E_c(0)=0,\quad I_c(0)=1,\quad
S_b(0)=499,\quad I_b(0)=1,\quad B(0)=0.
\]
These values represent scaled population levels chosen for numerical convenience and clarity of presentation. They do not correspond to specific herds, farms, or geographical regions. Numerical experiments conducted using alternative initial conditions yielded qualitatively similar behaviour, indicating that the results are robust to the choice of initial population sizes.

Model parameters used in the simulations were assigned uncertainty distributions representing plausible epidemiological variability. Representative baseline values used for the illustrative single-trajectory simulations are reported in Table~\ref{tab:parameters}. These baseline values are distinct from the random parameter sets generated from the
corresponding uncertainty distributions for the simulation ensembles. 

\begin{table}
\centering
\caption{Baseline parameter values, uncertainty distributions, and units used for numerical simulations of model~\eqref{eqn:reduced}}
\label{tab:parameters}
\resizebox{\textwidth}{!}{%
\begin{tabular}{|l|l|l|l|}
\toprule
\textbf{Parameter} 
& \textbf{Baseline value} 
& \textbf{Uncertainty distribution} 
& \textbf{Units / Source} \\
\midrule

$\Lambda_c$ 
& $30$ 
& Fixed 
& cattle day$^{-1}$ \\

$\Lambda_b$ 
& $10$ 
& Fixed 
& birds day$^{-1}$ \\

$\beta_c$ 
& $3.0\times10^{-3}$ 
& $\mathrm{LogNormal}$ on $[1.5\times10^{-3},\,6.0\times10^{-3}]$ 
& cattle$^{-1}$ day$^{-1}$ \\

$\beta_b$ 
& $4.0\times10^{-7}$ 
& $\mathrm{LogNormal}$ on $[2.0\times10^{-7},\,8.0\times10^{-7}]$ 
& bird$^{-1}$ day$^{-1}$~\cite{gumel2009}\\

$\beta_{bc}$ 
& $9.16\times10^{-6}$ 
& $\mathrm{LogNormal}$ on $[4.58\times10^{-6},\,1.83\times10^{-5}]$ 
& bird$^{-1}$ day$^{-1}$~\cite{CHANG2026} \\

$\beta_{\mathrm{env}}$ 
& $1.6\times10^{-5}$ 
& $\mathrm{LogNormal}$ on $[8.0\times10^{-6},\,3.2\times10^{-5}]$ 
& viral-unit$^{-1}$ day$^{-1}$\\

$\sigma_c$ 
& $0.51$ 
& $\mathrm{Gamma}$ on $[0.255,\,1.02]$ 
& day$^{-1}$~\cite{Rawson2025, PenaMosca2025}\\

$\gamma_c$ 
& $0.98$ 
& $\mathrm{Gamma}$ on $[0.49,\,1.96]$ 
& day$^{-1}$\\

$\gamma_b$ 
& $1.7\times10^{-4}$ 
& $\mathrm{Gamma}$ on $[8.5\times10^{-5},\,3.4\times10^{-4}]$ 
& day$^{-1}$ \\

$\theta_c$ 
& $0.29$ 
& Uniform on $[0.10,\,0.50]$ 
& viral units cattle$^{-1}$ day$^{-1}$\\

$\theta_b$ 
& $0.17$ 
& Uniform on $[0.05,\,0.30]$ 
& viral units bird$^{-1}$ day$^{-1}$\\

$\mu_{\mathrm{env}}$ 
& $0.08$ 
& Uniform on $[0.01,\,0.20]$  
& day$^{-1}$~\cite{Martin2018}\\

$\mu_c$ 
& $0.0002$ 
& Fixed 
& day$^{-1}$~\cite{AVMA2025}\\

$\mu_b$ 
& $0.02$ 
& Fixed 
& day$^{-1}$~\cite{gumel2009}\\

$\mu_{dc}$ 
& $0.02$ 
& Fixed 
& day$^{-1}$~\cite{DeVries2020}\\

$\mu_{db}$ 
& $0.02$ 
& Fixed 
& day$^{-1}$\\

\bottomrule
\end{tabular}%
}
\end{table}

The environmental compartment \(B(t)\) is measured in arbitrary viral concentration units representing relative environmental contamination. Accordingly, \(\beta_{\mathrm{env}}\), \(\theta_c\), and \(\theta_b\) are effective rates defined relative to this lumped environmental scale, while direct transmission rates are expressed per infectious host of the corresponding type.

\noindent\textit{Distributional assumptions.}
The choice of uncertainty distributions was based on the mathematical properties and epidemiological interpretation of each parameter class. Transmission parameters were assigned log-normal distributions because they are strictly positive and may vary over orders of magnitude. Progression and recovery rates were assigned gamma distributions because they represent positive transition rates associated with waiting-time processes. Environmental contamination and decay parameters were assigned uniform distributions to reflect limited prior knowledge and to avoid imposing unwarranted distributional structure. Fixed demographic and mortality parameters were retained at baseline values to focus the uncertainty analysis on transmission, recovery, and environmental persistence. \myEdit{
The two values reported for each uncertain parameter in Table~\ref{tab:parameters} denote the
lower and upper limits of the corresponding sampling interval, respectively,
rather than the defining parameters of the probability distribution. For the
simulation ensembles, samples were generated from the indicated distributional
family within these limits. These simulation distributions are distinct from
the uniform sampling scheme used in the main LHS--PRCC analysis.
} 

\subsection{Dynamics of the state variables of model \ref{eqn:reduced}}

\begin{figure}[ht!]
    \centering
    \includegraphics[width=1\linewidth]{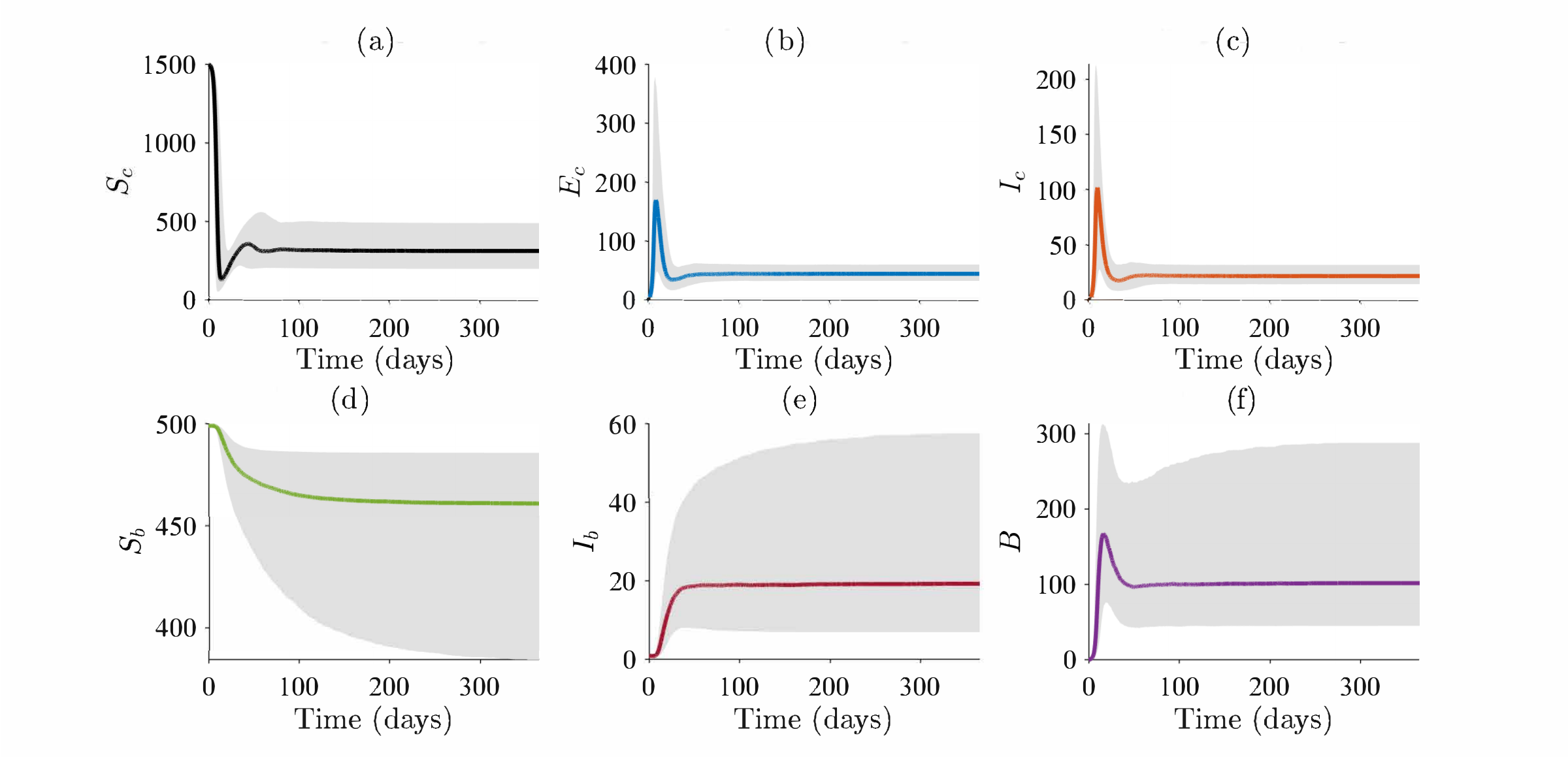}
    \caption{Simulated transmission dynamics of HPAI in cattle, wild birds, and the environment over 365 days with parameters as in Table~\ref{tab:parameters}}
    \label{fig:cattlesimul}
\end{figure}

Figure~\ref{fig:cattlesimul} presents the simulated transmission dynamics of HPAI in cattle, wild birds, and the contaminated environment over a period of 365 days. Panels~(a)--(c) show the dynamics of the cattle population. In panel~(a), the susceptible cattle population decreases rapidly from its initial value of approximately $1499$ cattle during the early phase of the outbreak, before gradually approaching a lower steady level. This decline reflects the progressive depletion of susceptible cattle as transmission occurs within the cattle population under the homogeneous-mixing assumption.

Panel~(b) shows that the exposed cattle population increases sharply shortly after the start of the simulation, reaching a peak of about $180$ cattle within the first few weeks. Thereafter, the exposed population declines steadily as exposed cattle progress to the infectious class. Similarly, the infected cattle population in panel~(c) rises rapidly following the increase in the exposed class, attaining a peak of approximately $100$ infected cattle during the early outbreak phase. The infected population subsequently decreases and approaches a very low level as infected cattle recover, die, or leave the infectious class.

Panels~(d)--(f) describe the corresponding dynamics in wild birds and the environment. The susceptible wild bird population, shown in panel~(d), remains relatively stable throughout the simulation, with only a slight decline from its initial level. This suggests that, under the chosen parameter values, infection has a comparatively limited effect on the total susceptible bird population. In panel~(e), the infected bird population initially changes slowly, then increases to a moderate peak before declining and settling at a persistent endemic level. This indicates sustained circulation of HPAI among wild birds over the simulation period. 

Finally, panel~(f) shows the concentration of HPAI virus in the environment. The environmental viral load increases rapidly during the early stage of the outbreak, reaching a peak of approximately $170$ units, before gradually declining as viral decay and reduced shedding dominate. The persistence of environmental contamination highlights the potential role of the environment as an indirect transmission pathway for maintaining HPAI infection in both cattle and wild bird populations.

\subsection{Extinction and persistence of infection}

To illustrate the threshold behaviour of system~\eqref{eqn:reduced}, we simulated the model under parameter regimes satisfying $\mathcal R_0<1$ and $\mathcal R_0>1$.

\begin{figure}[ht!]
    \centering
    \includegraphics[width=\textwidth]{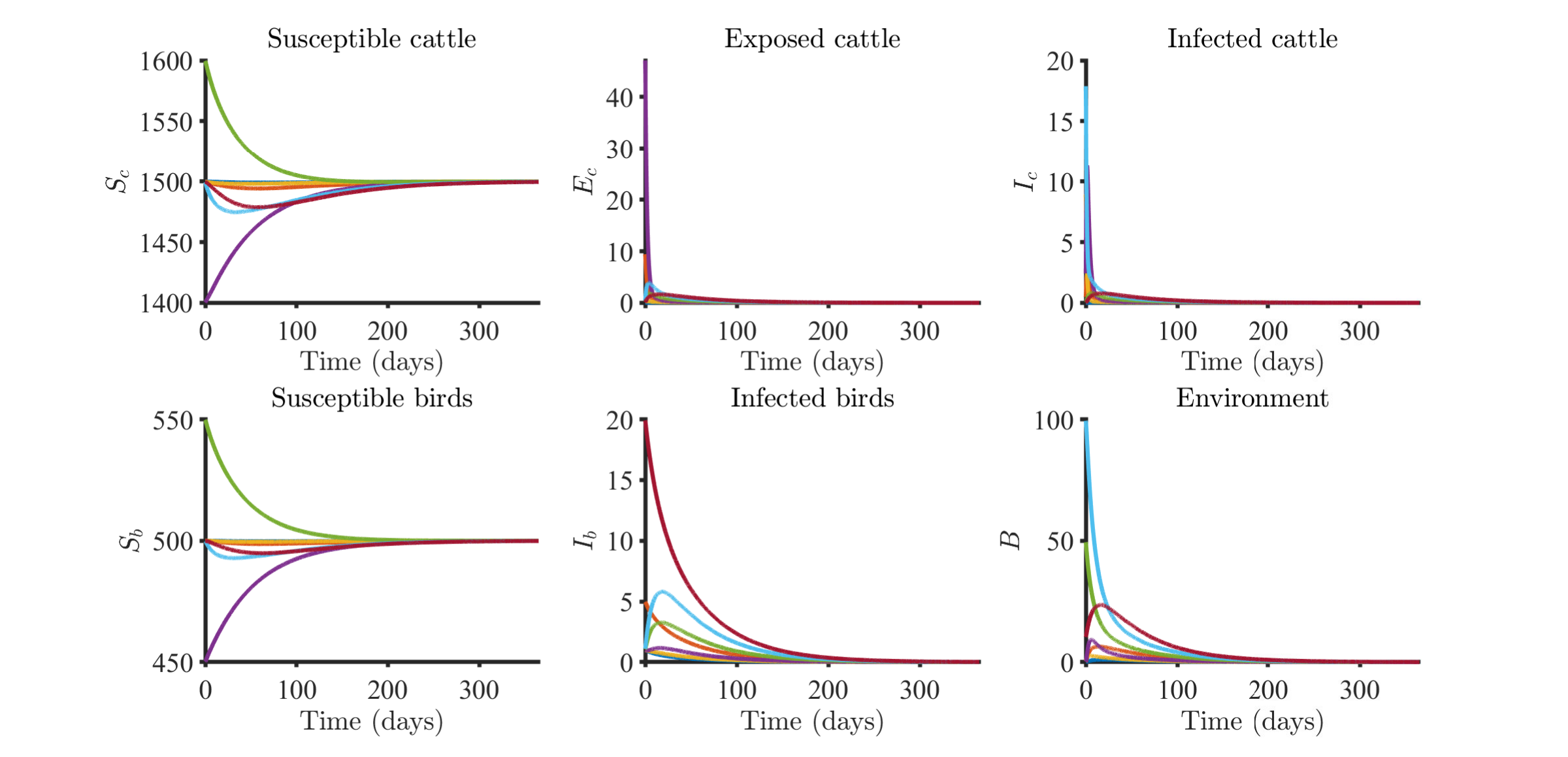}
    \caption{Convergence of trajectories initiated from different
    initial conditions to the disease-free equilibrium $\psi_{\text{DFE}}$
    when $\mathcal R_0<1$.}
    \label{fig:DFE_GS}
\end{figure}

Figure~\ref{fig:DFE_GS} shows that, when $\mathcal R_0<1$,
trajectories initiated from different initial conditions converge to
the disease-free equilibrium $\psi_{\text{DFE}}$. This numerical behaviour
is consistent with the global stability result established in
Theorem~\ref{thm:threshold_stability}. Thus, in the absence of external
viral introduction, infection is eliminated from the modelled system.

\begin{figure}[ht!]
    \centering
    \includegraphics[width=\textwidth]{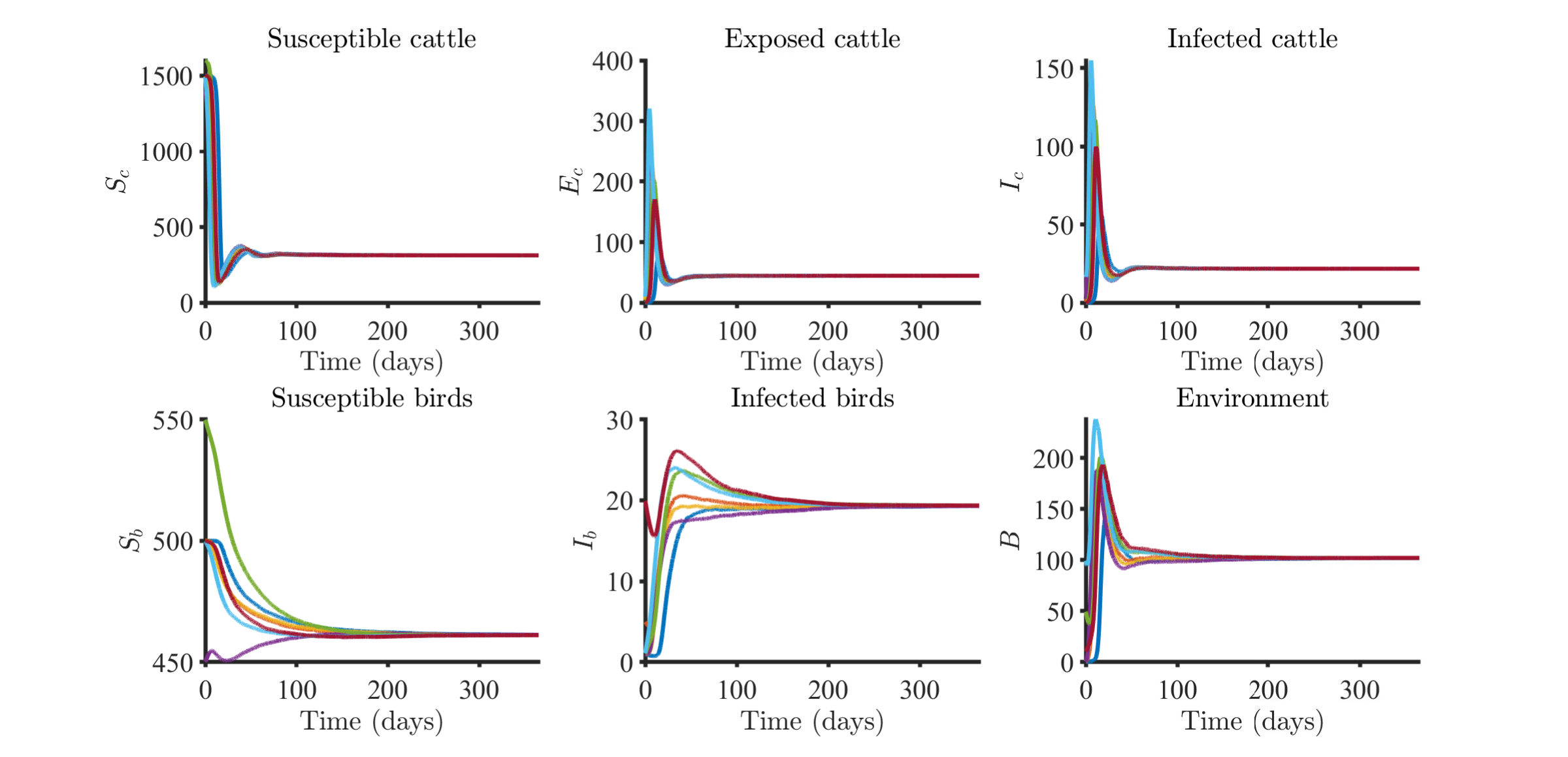}
    \caption{Convergence of trajectories initiated from different
    initial conditions to the positive endemic equilibrium
    $\psi_{\text{EE}}$ when $\mathcal R_0>1$.}
    \label{fig:EE_GS}
\end{figure}

Figure~\ref{fig:EE_GS} shows that, when $\mathcal R_0>1$, trajectories
initiated from different initial conditions converge to the unique
strictly positive endemic equilibrium $\psi_{\text{EE}}$ for the
parameter regime examined. This numerical behaviour is consistent with
the endemic-equilibrium stability result in
Theorem~\ref{thm:threshold_stability}. The simulations therefore
illustrate the persistence of infection at a positive endemic level
under the selected parameter regime.

Overall, the numerical results are consistent with the analytical
threshold behaviour of the closed model. When $\mathcal R_0<1$,
trajectories converge to the disease-free equilibrium and infection is
eliminated. When $\mathcal R_0>1$ and a unique strictly positive
endemic equilibrium exists under the assumptions of
Theorem~\ref{thm:threshold_stability}, trajectories converge to the
endemic equilibrium for the parameter regimes examined.

\subsection{Global sensitivity analysis}\label{sec:glosens}
\begin{figure*}[ht!]
    \centering
    \includegraphics[width=0.85\linewidth]{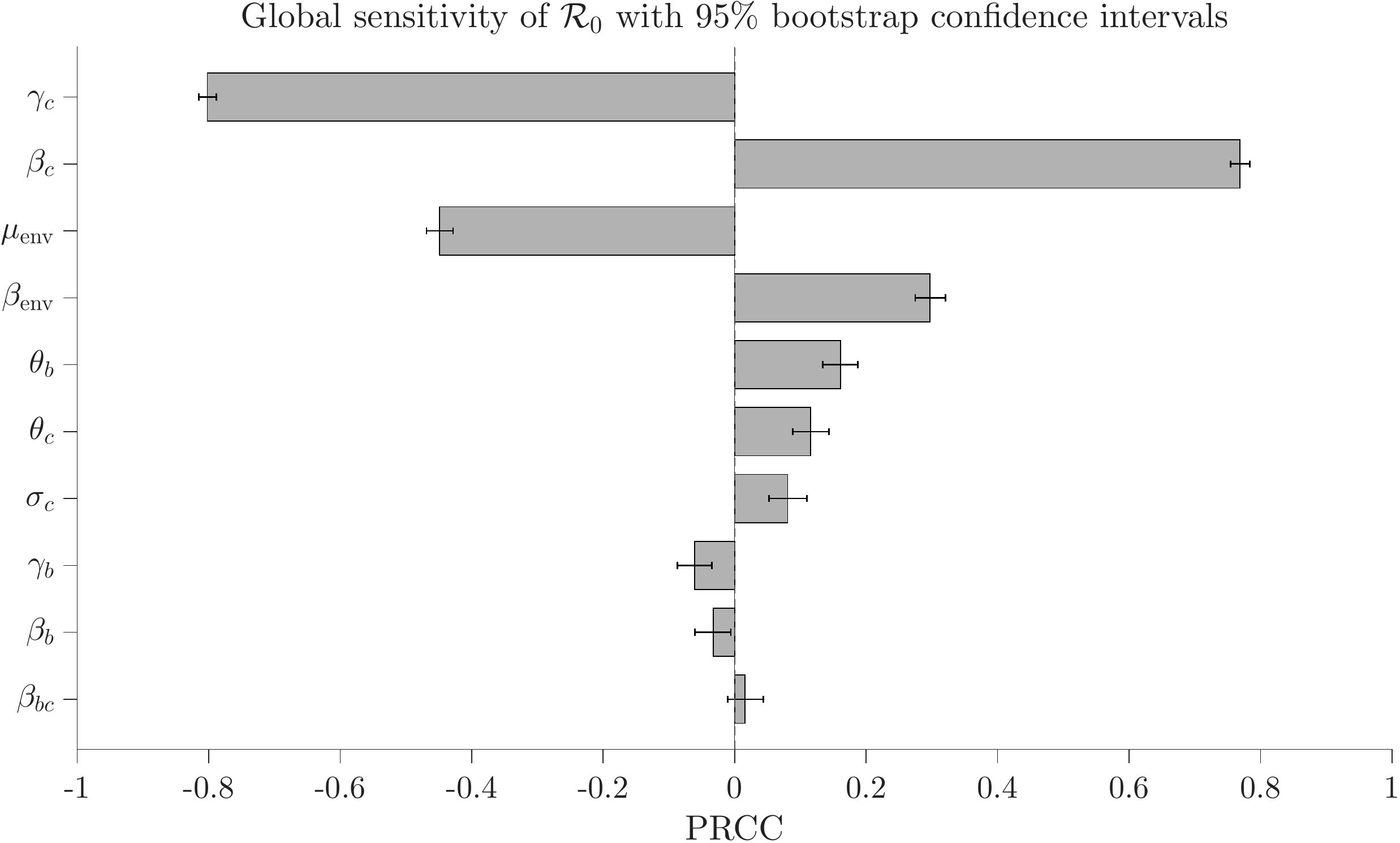}
    \caption{Global sensitivity analysis of the basic reproduction number $\mathcal{R}_0$ with respect to selected parameters of model~\eqref{eqn:reduced}. Bars show Partial Rank Correlation Coefficients (PRCC), while horizontal confidence bands indicate 95\% bootstrap confidence intervals}
    \label{fig:gloSens}
\end{figure*}

In this section, Latin Hypercube Sampling (LHS) combined with Partial Rank Correlation Coefficient (PRCC) analysis was used to quantify the influence of selected model parameters on the basic reproduction number $\mathcal{R}_0$. A total of $N=5000$ parameter sets were generated. The LHS--PRCC analysis used uniform sampling over the corresponding parameter ranges to ensure even coverage of the uncertainty space, whereas the simulation ensembles used the uncertainty distributions reported in Table~\ref{tab:parameters}. For each sampled parameter set, $\mathcal{R}_0$ was computed analytically using Eq.~\eqref{eq:R0}. Uncertainty in the PRCC estimates was quantified using bootstrap resampling, yielding 95\% confidence intervals and significance levels. The resulting PRCC values are shown in Fig.~\ref{fig:gloSens} and summarised in Table~\ref{tab:PRCC}.

Figure~\ref{fig:gloSens} illustrates both the magnitude and direction of the monotonic association between each parameter and $\mathcal{R}_0$. Positive PRCC values indicate parameters whose increase raises the transmission potential, whereas negative PRCC values indicate parameters whose increase reduces $\mathcal{R}_0$. Parameters whose confidence intervals do not include zero are interpreted as having statistically significant effects on the threshold quantity.

\begin{table}[ht!]
\centering
\caption{Partial rank correlation coefficients for the basic
reproduction number $\mathcal R_0$, with 95\% bootstrap confidence
intervals and significance levels.}
\label{tab:PRCC}
\begin{tabular}{lccccc}
\toprule
\textbf{Parameter}
&
\textbf{PRCC}
&
\textbf{CI 2.5\%}
&
\textbf{CI 97.5\%}
&
\(\boldsymbol{p}\)\textbf{-value}
&
\textbf{Significance}
\\
\midrule

$\beta_c$
& 0.7688
& 0.7518
& 0.7849
& $<0.0001$
& *** \\

$\beta_b$
& $-0.0322$
& $-0.0605$
& $-0.0039$
& 0.0230
& * \\

$\beta_{bc}$
& 0.0159
& $-0.0119$
& 0.0434
& 0.2602
& -- \\

$\beta_{\mathrm{env}}$
& 0.2972
& 0.2736
& 0.3209
& $<0.0001$
& *** \\

$\sigma_c$
& 0.0807
& 0.0529
& 0.1091
& $<0.0001$
& *** \\

$\gamma_c$
& $-0.8022$
& $-0.8161$
& $-0.7877$
& $<0.0001$
& *** \\

$\gamma_b$
& $-0.0606$
& $-0.0875$
& $-0.0330$
& $<0.0001$
& *** \\

$\theta_c$
& 0.1156
& 0.0875
& 0.1436
& $<0.0001$
& *** \\

$\theta_b$
& 0.1612
& 0.1343
& 0.1874
& $<0.0001$
& *** \\

$\mu_{\mathrm{env}}$
& $-0.4488$
& $-0.4693$
& $-0.4262$
& $<0.0001$
& *** \\

\bottomrule
\end{tabular}

\begin{minipage}{0.92\textwidth}
\footnotesize
\textit{Notes:} PRCC values quantify the adjusted monotonic
associations between the ranked model parameters and ranked values of
$\mathcal R_0$ after controlling for the ranks of the remaining
parameters. Confidence intervals were obtained by nonparametric
bootstrap resampling, with the rank transformation and residualisation
repeated within each bootstrap replicate. Two-sided $p$-values were
calculated using the partial-correlation $t$-test with 4989 degrees of
freedom. Values smaller than $0.0001$ are reported as
$p<0.0001$. Significance levels are denoted by
*** \(p<0.001\), ** \(p<0.01\), and * \(p<0.05\);
``--'' indicates non-significance at the 5\% level.
\end{minipage}
\end{table}

The results show that $\mathcal R_0$ is most strongly governed by
cattle recovery and cattle-to-cattle transmission. The cattle
transmission rate $\beta_c$ has a strong positive PRCC value of
$0.7688$, indicating that increases in effective direct cattle
transmission substantially increase $\mathcal R_0$. In contrast, the
cattle recovery rate $\gamma_c$ has the strongest negative association
with $\mathcal R_0$ (PRCC $=-0.8022$), showing that faster recovery or
removal of infected cattle reduces transmission potential. These two
parameters therefore form the dominant control axis of the system
within the parameter ranges considered.

Environmental processes also have important effects. The environmental
transmission coefficient $\beta_{\mathrm{env}}$ is positively
associated with $\mathcal R_0$ (PRCC $=0.2972$), whereas the
environmental decay or removal rate $\mu_{\mathrm{env}}$ has a
substantial negative association (PRCC $=-0.4488$). These results
indicate that environmentally mediated transmission can increase
transmission potential, while faster viral decay or removal from the
environment reduces $\mathcal R_0$. The environmental contamination
rates from infected cattle and wild birds, $\theta_c$ and $\theta_b$,
also have positive associations with $\mathcal R_0$, with PRCC values
of $0.1156$ and $0.1612$, respectively. Their effects are weaker than
those of $\beta_c$, $\gamma_c$, $\mu_{\mathrm{env}}$, and
$\beta_{\mathrm{env}}$.

The wild-bird recovery rate $\gamma_b$ has a weak negative but
statistically significant association with $\mathcal R_0$
(PRCC $=-0.0606$). This indicates that increasing the
recovery rate of infected wild birds reduces their contribution to
direct and environmentally mediated transmission. The direct
bird-to-bird transmission rate $\beta_b$ also has a small negative
association with $\mathcal R_0$ (PRCC $=-0.0322$).
Although this association is statistically significant, its magnitude
is small, indicating a limited influence within the parameter ranges
considered.

The corrected spectral-radius formulation of $\mathcal R_0$ explicitly
includes the cross-species transmission coefficient $\beta_{bc}$.
However, $\beta_{bc}$ has a small positive PRCC value of $0.0159$, and
its 95\% confidence interval, from $-0.0104$ to $0.0439$, includes
zero. Its association with $\mathcal R_0$ is therefore not
statistically significant under the principal sampling scheme
($p=0.2378$). This result does not imply that the cross-species pathway
is absent from the model; rather, it indicates that variation in
$\beta_{bc}$ has a comparatively weak independent association with
$\mathcal R_0$ over the parameter range examined after controlling for
the remaining parameters.

The cattle progression rate $\sigma_c$ has a positive and statistically
significant association with $\mathcal R_0$ (PRCC $=0.0807$;
$p=0.0020$). This reflects the fact that faster progression from the
exposed class to the infectious class increases the proportion of
infected cattle that survive the latent stage and subsequently
contribute to direct transmission and environmental contamination.
Nevertheless, the influence of $\sigma_c$ remains weaker than those of
the dominant cattle transmission, cattle recovery, and environmental
parameters.

\myEdit{
To assess whether the sensitivity conclusions depend on the assumed parameter
distributions and sampling ranges, the LHS--PRCC analysis was repeated under
two alternative sampling schemes. In addition to uniform sampling over the
original sensitivity-analysis ranges, we considered log-uniform sampling over
the same ranges and uniform sampling over ranges widened by 50\%. For each
scheme, 5000 parameter sets were generated, and 95\% confidence intervals were
obtained using 1000 bootstrap replicates. The resulting PRCC estimates are
compared in Fig.~\ref{fig:PRCC_robustness}.

\begin{figure}[ht]
\centering
\includegraphics[width=\textwidth]{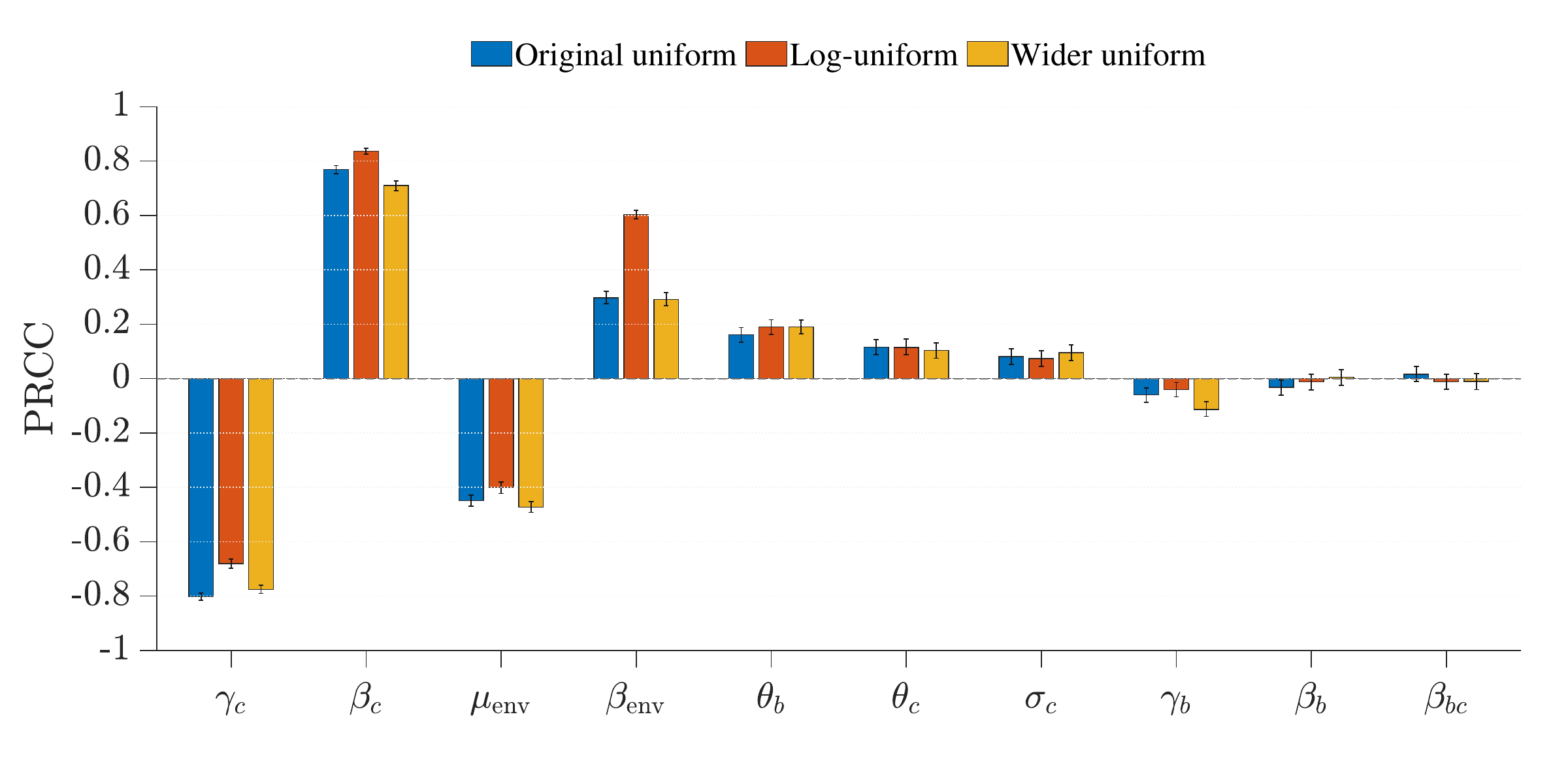}
\caption{Robustness of the PRCC estimates for the basic reproduction number
$\mathcal{R}_0$ under alternative parameter-sampling schemes. The original
analysis uses uniform sampling over the prescribed sensitivity-analysis
ranges. The alternative analyses use log-uniform sampling over the same ranges
and uniform sampling over ranges widened by 50\%. Bars represent the PRCC
estimates, and error bars indicate 95\% bootstrap confidence intervals.
Parameters are ordered according to the absolute magnitudes of their PRCC
estimates under the original uniform scheme.}
\label{fig:PRCC_robustness}
\end{figure}

As shown in Fig.~\ref{fig:PRCC_robustness}, the directions and broad
relative importance of the principal parameter effects are generally
consistent across the three sampling schemes. The cattle recovery rate
$\gamma_c$ remains the strongest negative determinant of
$\mathcal R_0$, whereas the cattle-to-cattle transmission rate
$\beta_c$ remains the strongest positive determinant. These two
parameters occupy the leading positions in the absolute PRCC ranking
under the alternative uncertainty schemes, although their precise
magnitudes vary with the assumed sampling distribution and parameter
ranges.

The environmental decay rate $\mu_{\mathrm{env}}$ also retains a
substantial negative association with $\mathcal R_0$. Thus, increasing
the rate at which infectious virus is removed from the shared
environment consistently reduces transmission potential. Conversely,
the environmental transmission coefficient
$\beta_{\mathrm{env}}$ remains positively associated with
$\mathcal R_0$. Its PRCC magnitude varies more noticeably among the
three sampling schemes, indicating that the estimated importance of
environmentally mediated transmission is comparatively sensitive to
the assumed uncertainty structure.

The wild-bird environmental contamination rate $\theta_b$ retains a
positive association with $\mathcal R_0$, indicating that increased
viral contamination by infected wild birds raises transmission
potential. The wild-bird recovery rate $\gamma_b$ remains negatively
associated with $\mathcal R_0$, although its effect is weaker than
that of cattle recovery. The cattle progression rate $\sigma_c$ and
the cattle environmental contamination rate $\theta_c$ also show
positive associations, but their effects are smaller than those of
the dominant cattle transmission, cattle recovery, and environmental
parameters.

The PRCC results indicate that $\beta_{bc}$ is positively associated
with $\mathcal R_0$, as expected biologically, because increases in
direct wild-bird-to-cattle transmission increase the invasion
potential of infection. However, its influence is comparatively small
over the parameter range examined. The direct bird-to-bird
transmission coefficient $\beta_b$ also has a relatively weak
association with $\mathcal R_0$ compared with the principal cattle and
environmental parameters.

Overall, the robustness analysis indicates that the principal
conclusions are not determined by a single choice of sampling
distribution or parameter range. In particular, cattle recovery,
cattle-to-cattle transmission, environmental viral removal, and
environmental transmission remain the most influential determinants
of $\mathcal R_0$ across the uncertainty schemes considered.
Nevertheless, the precise PRCC magnitudes and the ordering of the
lower-ranked parameters vary among the schemes. Robustness is
therefore interpreted as consistency in the direction and broad
relative importance of the principal effects, rather than complete
invariance of the numerical PRCC estimates or parameter rankings.
}

Thus, the PRCC results indicate that, \myEdit{within the parameter ranges
and modelling assumptions considered here, reducing cattle-to-cattle
transmission and increasing cattle recovery or removal appear to provide the
greatest reductions in transmission potential, while environmental management
may serve as a complementary strategy.} These findings motivate the
intervention scenarios considered in the following section, in which the
parameters identified as most influential by the sensitivity analysis are
varied to assess their relative effects on model outcomes.

\begin{figure*}[ht!]
\centering
 \begin{subfigure}[b]{.6\linewidth}
   \centering
   \caption{}
   \includegraphics[width=\textwidth]{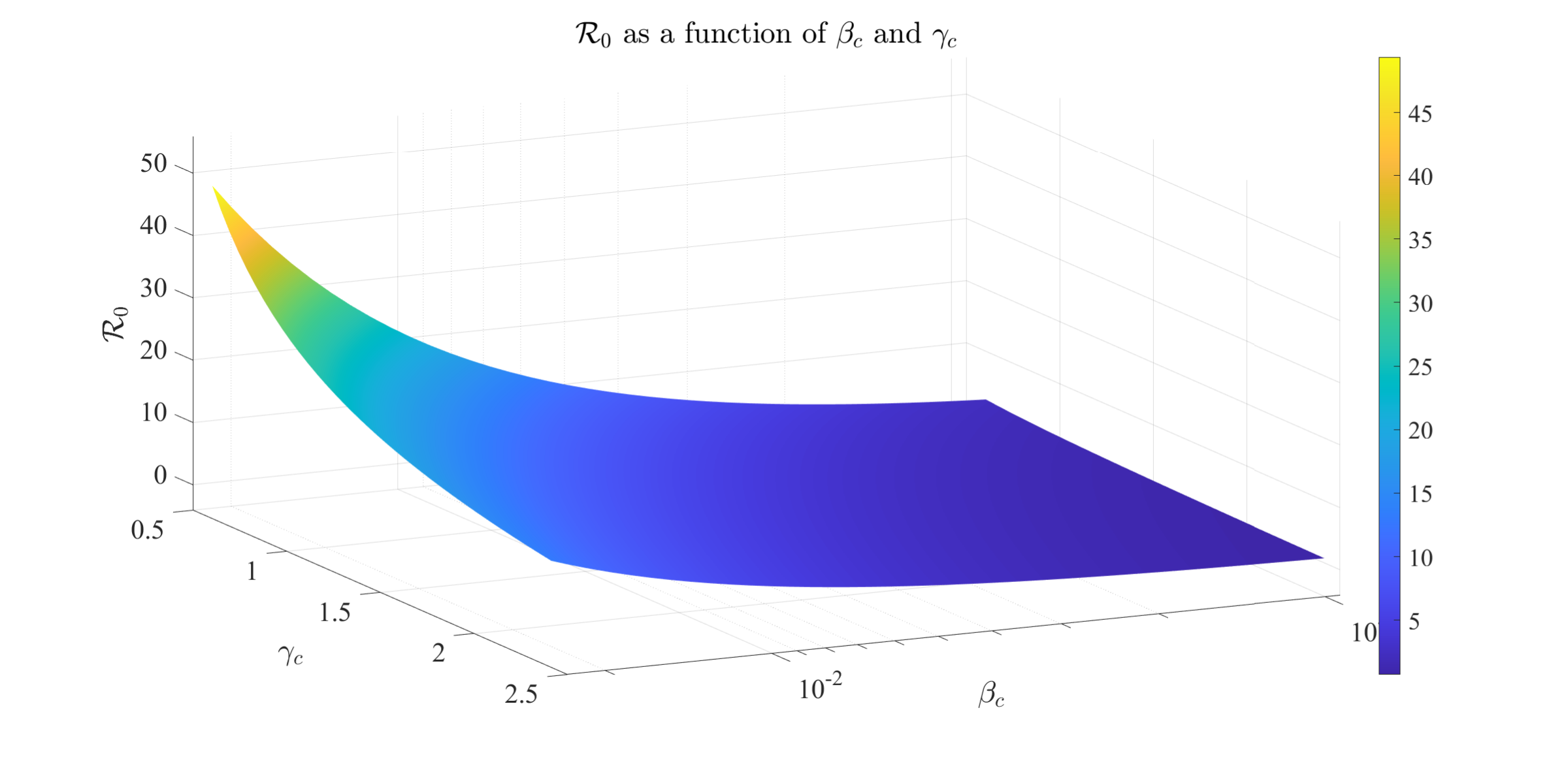}
   \label{fig:betaCgammaC}
 \end{subfigure}%
 \begin{subfigure}[b]{.55\linewidth}
   \centering
   \caption{}
   \includegraphics[width=\textwidth]{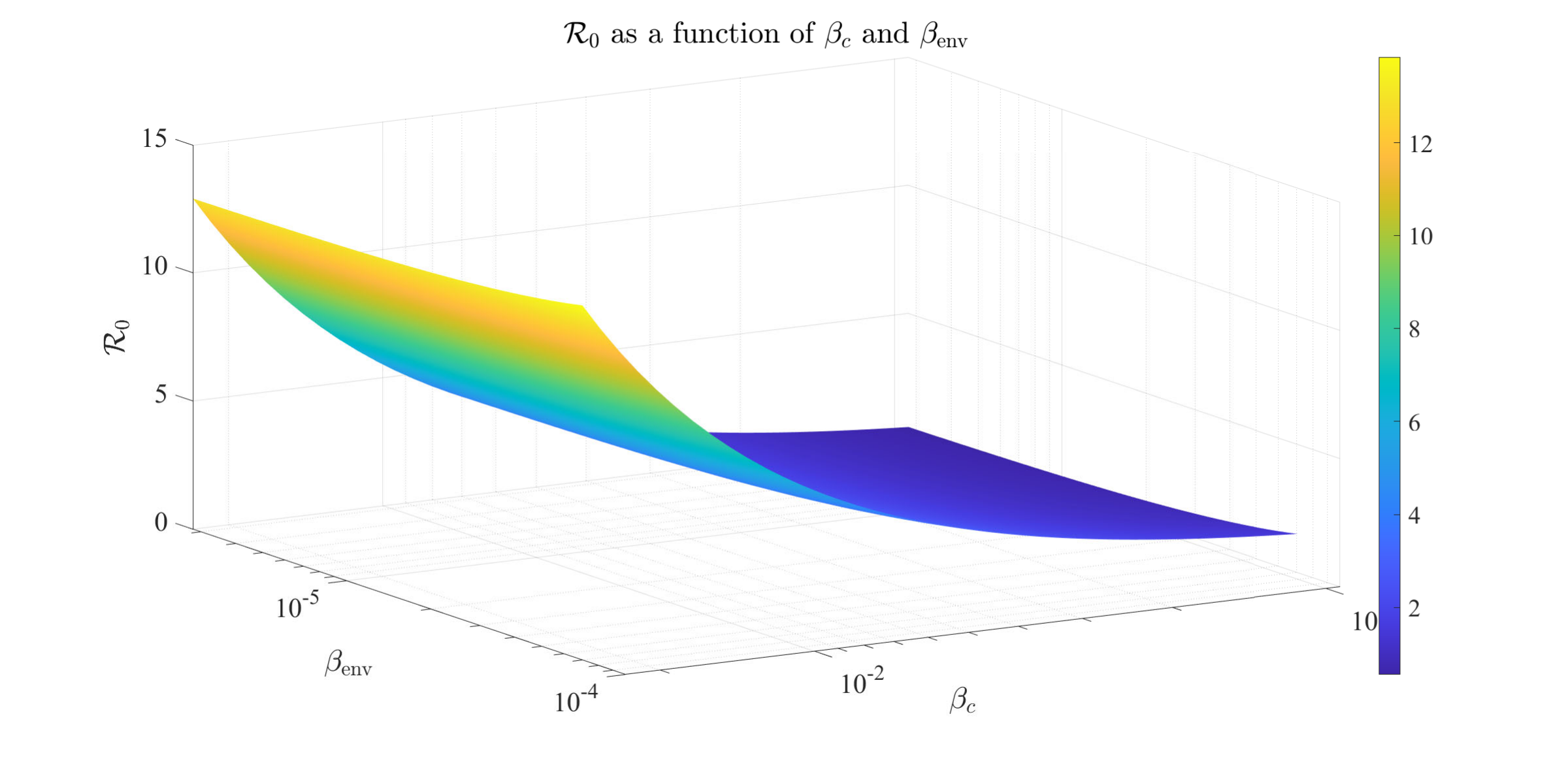}
   \label{fig:betaCbetaEnv}
 \end{subfigure}\\%
 \begin{subfigure}[b]{.6\linewidth}
   \centering
   \caption{}
   \includegraphics[width=\textwidth]{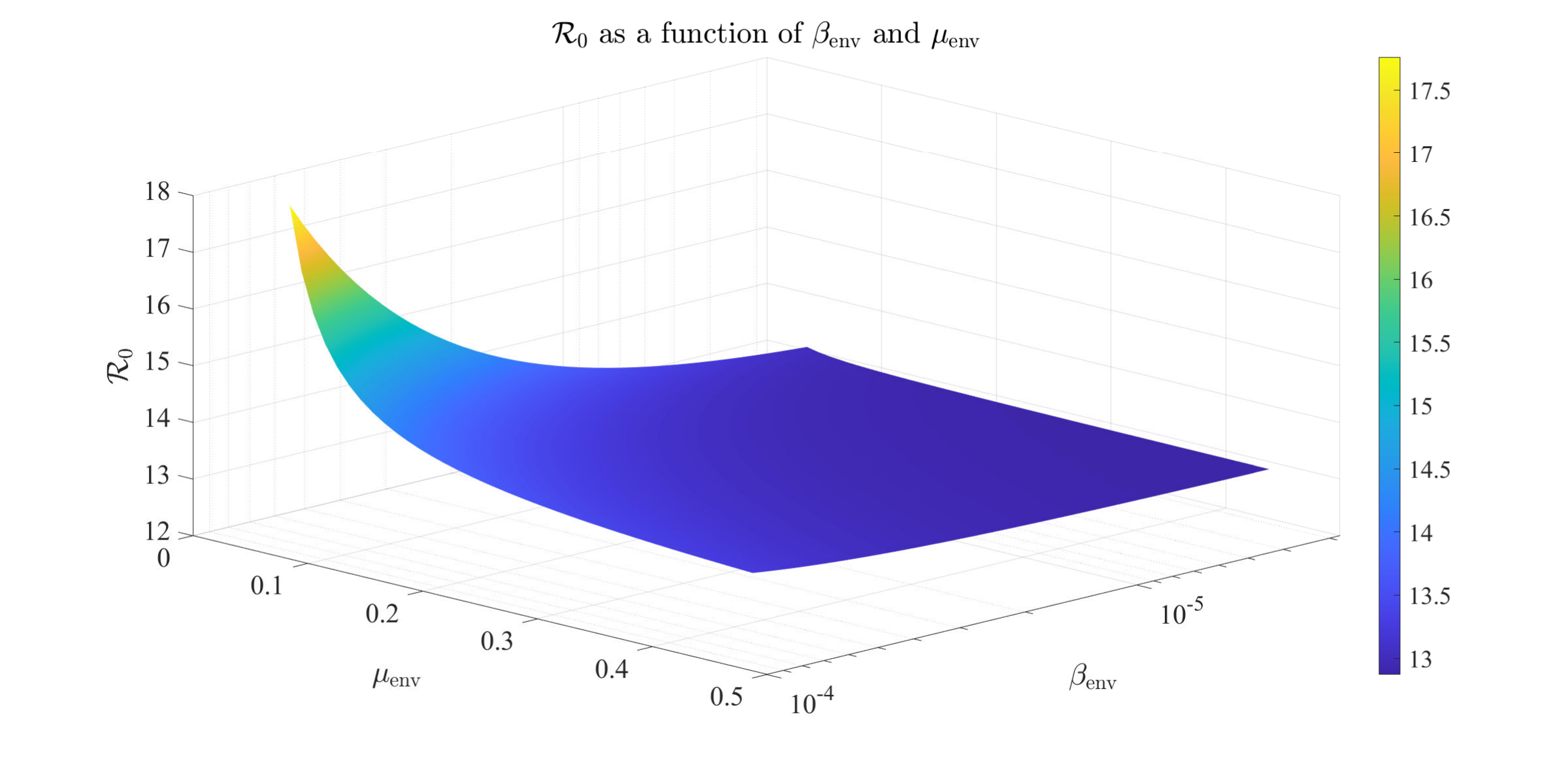}
   \label{fig:betaEnvmuEnv}
 \end{subfigure}%
 \caption{Surface plots of the basic reproduction number $\mathcal{R}_0$ as a function of selected pairs of influential model parameters: 
 (a) cattle--cattle transmission rate $\beta_c$ and cattle recovery rate $\gamma_c$; 
 (b) cattle--cattle transmission rate $\beta_c$ and environmental transmission rate $\beta_{\mathrm{env}}$; 
 (c) environmental transmission rate $\beta_{\mathrm{env}}$ and environmental decay rate $\mu_{\mathrm{env}}$}
 \label{fig:contourplots}
\end{figure*}

Next, we examined the joint influence of pairs of key model parameters on the basic reproduction number $\mathcal{R}_{0}$ in order to further characterise interaction effects and nonlinear threshold behaviour. Fig.~\ref{fig:contourplots} presents three surface plots showing how $\mathcal{R}_{0}$ varies as a function of selected parameter pairs identified as influential in the global sensitivity analysis. In each plot, the colour scale represents the magnitude of $\mathcal{R}_{0}$, with values exceeding unity corresponding to epidemic invasion.

Figure~\ref{fig:betaCgammaC} shows the dependence of $\mathcal{R}_{0}$ on the cattle--cattle transmission rate $\beta_c$ and the cattle recovery rate $\gamma_c$. The surface exhibits a strongly monotonic structure: $\mathcal{R}_{0}$ increases rapidly as $\beta_c$ increases and decreases as $\gamma_c$ increases. For sufficiently low values of $\beta_c$ and high values of $\gamma_c$, the system remains below the epidemic threshold $(\mathcal{R}_{0}<1)$, indicating disease fade-out. However, moderate increases in $\beta_c$ or reductions in $\gamma_c$ are sufficient to push $\mathcal{R}_{0}$ above unity. This interaction highlights the dominant role of within-cattle transmission and recovery processes in shaping outbreak potential, and is consistent with the strong PRCC values obtained for $\beta_c$ and $\gamma_c$.

Figure~\ref{fig:betaCbetaEnv} illustrates the combined effects of cattle--cattle transmission $\beta_c$ and environmental transmission $\beta_{\mathrm{env}}$. The surface indicates that both parameters contribute positively to $\mathcal{R}_{0}$, with the largest values occurring when both direct cattle transmission and environmentally mediated transmission are high. Elevated values of $\beta_{\mathrm{env}}$ can increase $\mathcal{R}_{0}$ even when $\beta_c$ is moderate, suggesting that environmental transmission can amplify outbreaks primarily driven by cattle contact. Conversely, when $\beta_c$ is low, reductions in $\beta_{\mathrm{env}}$ move the system closer to the epidemic threshold, reinforcing the role of environmental transmission as an important supporting pathway.

The interaction between environmental transmission and pathogen decay is shown in Fig.~\ref{fig:betaEnvmuEnv}. The surface demonstrates that $\mathcal{R}_{0}$ increases with the environmental transmission rate $\beta_{\mathrm{env}}$ and decreases with the environmental decay rate $\mu_{\mathrm{env}}$. When pathogen clearance from the environment is slow, even moderate values of $\beta_{\mathrm{env}}$ can maintain $\mathcal{R}_{0}>1$, highlighting the importance of environmental persistence. In contrast, higher environmental decay rates substantially reduce transmission potential, even in the presence of environmental exposure.

Taken together, these surface plots complement the PRCC results by revealing how key parameters interact nonlinearly to determine the epidemic threshold. While effective cattle--cattle transmission and cattle recovery dominate the dynamics, environmental transmission and persistence act as amplifying mechanisms that can shift the system across the epidemic threshold. From a control perspective, these findings suggest that reducing cattle transmission or accelerating cattle recovery should be prioritised, while environmental management can provide additional benefits by limiting indirect transmission.

\subsection{Intervention scenarios and control effectiveness}
\label{sec:interventions}

To complement the global sensitivity analysis in
Section~\ref{sec:glosens}, we evaluated intervention scenarios
targeting parameters identified as influential determinants of the
basic reproduction number $\mathcal R_0$. Specifically, we considered
interventions acting on the cattle-to-cattle transmission rate
$\beta_c$, the cattle recovery rate $\gamma_c$, and the
environmentally mediated transmission coefficient
$\beta_{\mathrm{env}}$. These parameters represent three distinct
control mechanisms: reducing direct transmission among cattle,
increasing the effective recovery or removal rate of infected cattle,
and reducing indirect transmission through environmental exposure.

The intervention analysis used the fixed baseline parameter values
reported in Table~\ref{tab:parameters}. The same baseline parameter set and initial
conditions were used for every scenario, with only the targeted
parameter modified. Under this baseline parameterisation, the basic
reproduction number calculated from Equation~\eqref{eq:R0} was $\mathcal R_0=4.423.$

For the transmission-related parameters, moderate and strong
interventions were represented by reductions of $25\%$ and $75\%$,
respectively, in $\beta_c$ and $\beta_{\mathrm{env}}$. For the
recovery parameter $\gamma_c$, moderate and strong interventions were
represented by increases of $25\%$ and $75\%$, respectively. These
increases may represent improved detection, treatment, isolation, or
removal of infected cattle within the theoretical framework.

\begin{figure*}[ht!]
\centering
\includegraphics[width=\textwidth]{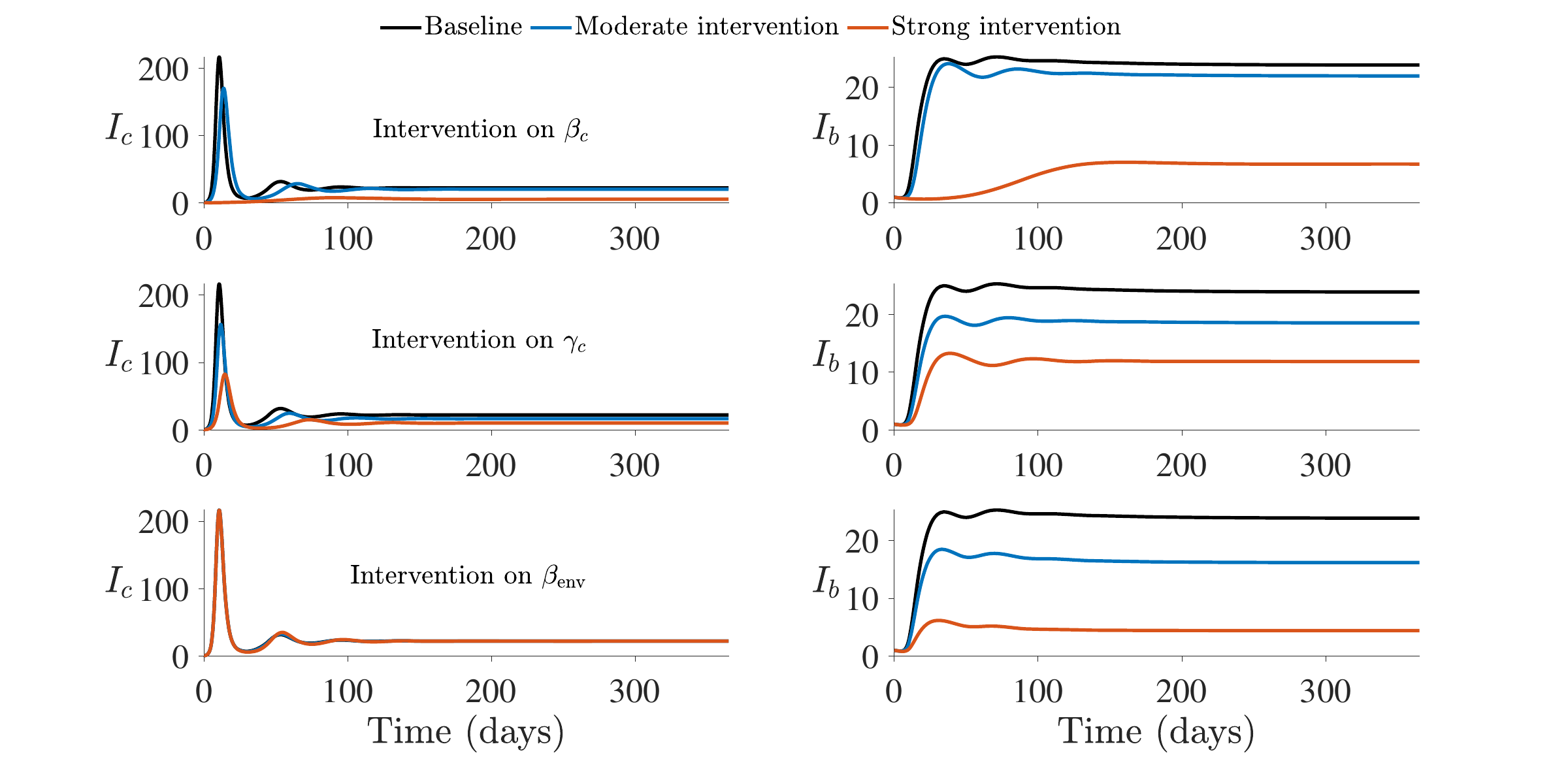}
\caption{Intervention scenario analysis using the fixed baseline
parameter values reported in Table~\ref{tab:parameters}. The scenarios compare the
baseline dynamics with moderate and strong interventions targeting
cattle-to-cattle transmission $\beta_c$, cattle recovery
$\gamma_c$, and environmentally mediated transmission
$\beta_{\mathrm{env}}$. Moderate and strong interventions correspond
to reductions of $25\%$ and $75\%$ in $\beta_c$ and
$\beta_{\mathrm{env}}$, and increases of $25\%$ and $75\%$ in
$\gamma_c$, respectively.}
\label{fig:intervention}
\end{figure*}

For each scenario, $\mathcal R_0$ was recalculated using
Equation~\eqref{eq:R0}, and the model was simulated over the specified
time horizon. We recorded the peak number of infected cattle, the
percentage reduction in the peak relative to the fixed baseline, and
the time to peak infection. Figure~\ref{fig:intervention} compares the
infected-cattle and infected-wild-bird trajectories under the
baseline, moderate-intervention, and strong-intervention scenarios.

The results show that interventions targeting $\beta_c$ produce the
strongest overall epidemiological effect. A $25\%$ reduction in
$\beta_c$ lowers $\mathcal R_0$ from $4.423$ to $3.345$,
corresponding to a reduction of approximately $24.37\%$. The peak
number of infected cattle decreases from approximately $217$ to
$170$, representing a reduction of approximately $21.36\%$.

A $75\%$ reduction in $\beta_c$ produces a substantially greater
effect, lowering $\mathcal R_0$ to $1.222$, corresponding to a
reduction of approximately $72.36\%$. The peak number of infected
cattle decreases to approximately $8$, representing a reduction
of approximately $96.35\%$. Although this intervention substantially
reduces both transmission potential and peak cattle infection,
$\mathcal R_0$ remains above unity. Thus, the strong reduction in
$\beta_c$ does not, by itself, reduce the system below the invasion
threshold under the baseline parameterisation.

Increasing the cattle recovery rate $\gamma_c$ also substantially
reduces transmission potential and peak cattle infection. A $25\%$
increase in $\gamma_c$ lowers $\mathcal R_0$ from $4.423$ to
$3.555$, corresponding to a reduction of approximately $19.62\%$.
The peak number of infected cattle decreases from approximately
$217$ to $157$, representing a reduction of approximately
$27.76\%$.

A $75\%$ increase in $\gamma_c$ lowers $\mathcal R_0$ to $2.555$,
corresponding to a reduction of approximately $42.23\%$, while the
peak number of infected cattle decreases to approximately $83$,
representing a reduction of approximately $61.67\%$. These findings
indicate that increasing the effective recovery or removal rate of
infected cattle can considerably reduce outbreak severity. Although
the reduction in $\mathcal R_0$ is smaller than that obtained from a
$75\%$ reduction in $\beta_c$, the recovery interventions produce
substantial reductions in the peak number of infected cattle.

In contrast, interventions targeting $\beta_{\mathrm{env}}$ alone
produce comparatively small changes in the cattle outbreak
trajectories. A $25\%$ reduction in
$\beta_{\mathrm{env}}$ lowers $\mathcal R_0$ from $4.423$ to
$4.397$, corresponding to a reduction of approximately $0.58\%$.
The associated reduction in the peak number of infected cattle is
approximately $0.04\%$.

A $75\%$ reduction in $\beta_{\mathrm{env}}$ lowers
$\mathcal R_0$ to $4.350$, corresponding to a reduction of
approximately $1.65\%$, while the peak number of infected cattle
decreases by approximately $0.13\%$. Under the baseline
parameterisation and intervention scenarios considered,
environmentally mediated transmission therefore acts primarily as an
amplifying pathway rather than as the principal driver of cattle
infection dynamics.

A summary of the intervention outcomes is provided in
Table~\ref{tab:intervention_results}. Within the model assumptions and
parameter values considered, reducing direct cattle-to-cattle
transmission produces the greatest combined reduction in
$\mathcal R_0$ and peak cattle infection. Increasing cattle recovery
or removal provides an additional effective control pathway. The
comparatively small effect of reducing $\beta_{\mathrm{env}}$ alone
suggests that environmental management may be more beneficial when
implemented alongside measures targeting cattle transmission or the
infectious period. These findings provide qualitative comparisons of
potential control mechanisms and may inform future policy discussions
when considered alongside empirical evidence, local epidemiological
conditions, and practical implementation constraints.

\begin{table}[ht!]
\centering
\caption{Effects of moderate and strong interventions on the basic
reproduction number and the peak number of infected cattle using the
fixed baseline parameter values reported in Table~\ref{tab:parameters}. Percentage
reductions are calculated relative to the fixed baseline scenario.}
\label{tab:intervention_results}
\begin{tabular}{lcccc}
\toprule
\textbf{Scenario}
&
\(\boldsymbol{\mathcal R_0}\)
&
\begin{tabular}[c]{@{}c@{}}
\(\boldsymbol{\mathcal R_0}\) \textbf{reduction}\\
\textbf{(\%)}
\end{tabular}
&
\begin{tabular}[c]{@{}c@{}}
\textbf{Peak}\\
\(\boldsymbol{I_c}\)
\end{tabular}
&
\begin{tabular}[c]{@{}c@{}}
\textbf{Peak \(\boldsymbol{I_c}\) reduction}\\
\textbf{(\%)}
\end{tabular}
\\
\midrule

\(\beta_c\): Baseline
& 4.423 & 0.00 & 217 & 0.00 \\

\(\beta_c\): 25\% reduction
& 3.345 & 24.37 & 170 & 21.36 \\

\(\beta_c\): 75\% reduction
& 1.222 & 72.36 & 8 & 96.35 \\

\midrule

\(\gamma_c\): Baseline
& 4.423 & 0.00 & 217 & 0.00 \\

\(\gamma_c\): 25\% increase
& 3.555 & 19.62 & 157 & 27.76 \\

\(\gamma_c\): 75\% increase
& 2.555 & 42.23 & 83 & 61.67 \\

\midrule

\(\beta_{\mathrm{env}}\): Baseline
& 4.423 & 0.00 & 217 & 0.00 \\

\(\beta_{\mathrm{env}}\): 25\% reduction
& 4.397 & 0.58 & 217 & 0.04 \\

\(\beta_{\mathrm{env}}\): 75\% reduction
& 4.350 & 1.65 & 216 & 0.13 \\

\bottomrule
\end{tabular}
\end{table}

\subsubsection{Joint control strategy}

The intervention scenarios considered above indicate that reducing cattle--cattle transmission and increasing cattle recovery are among the most effective strategies for reducing outbreak severity. To further examine whether simultaneous implementation of these controls provides additional benefit, we considered a joint control strategy acting on both the cattle transmission rate $\beta_c$ and the cattle recovery rate $\gamma_c$.

We introduced two constant control parameters, $u_1$ and $u_2$, where $u_1,u_2\in[0,1]$. The first control, $u_1$, represents interventions that reduce effective cattle--cattle transmission, such as improved biosecurity, movement restriction, segregation of infected animals, and reduced contact during management practices. The second control, $u_2$, represents interventions that increase the effective recovery or removal rate of infected cattle, such as faster detection, isolation, treatment, or culling. Under these controls, the parameters are modified as
\[
\beta_c \longrightarrow (1-u_1)\beta_c,
\qquad
\gamma_c \longrightarrow (1+u_2)\gamma_c .
\]
Thus, increasing $u_1$ reduces the effective cattle transmission rate, while increasing $u_2$ shortens the infectious period by increasing the recovery or removal rate.

\begin{figure}[ht!]
\centering
\includegraphics[width=0.75\linewidth]{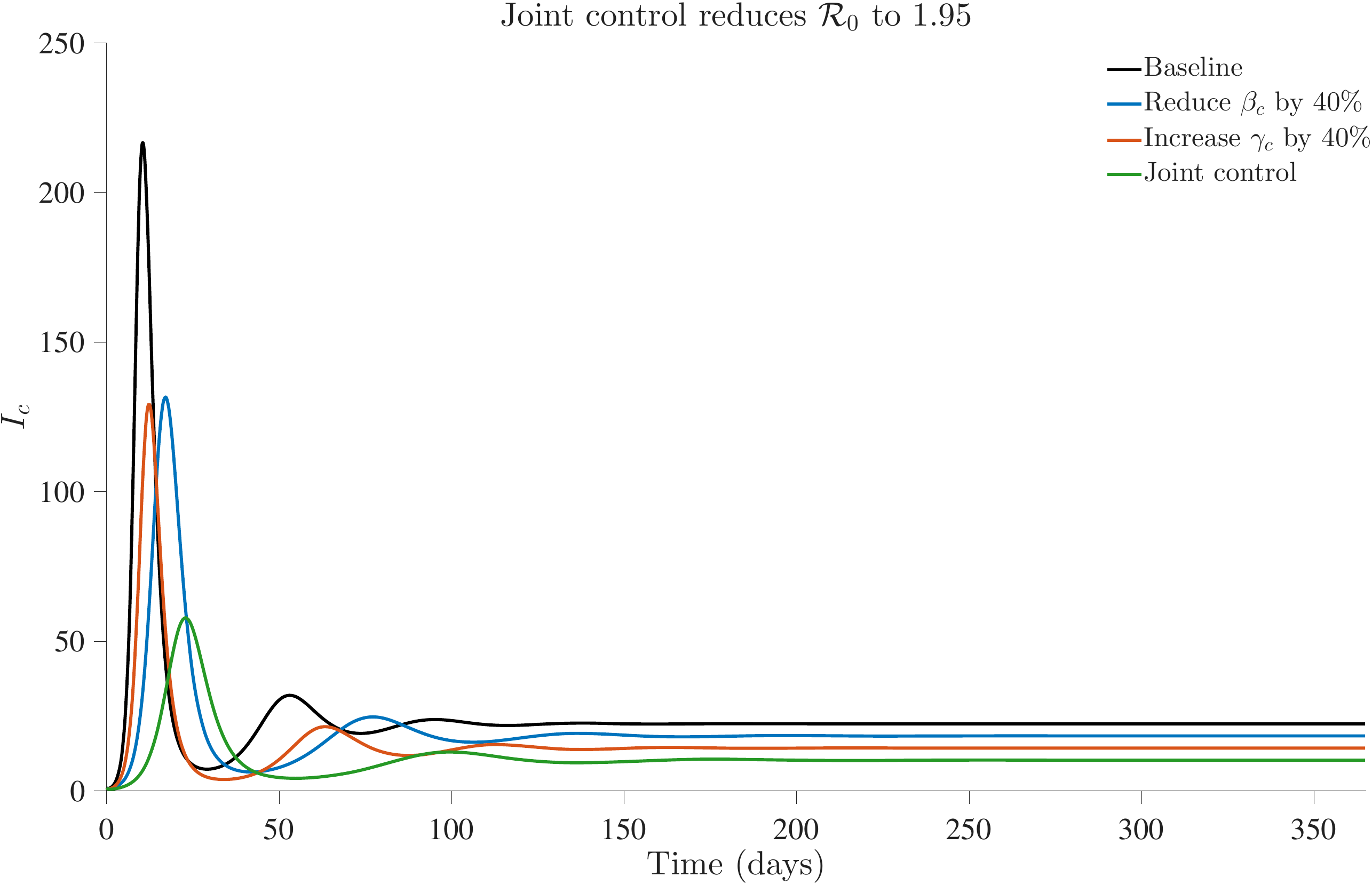}
\caption{Joint control analysis showing the effect of reducing cattle--cattle transmission $\beta_c$, increasing cattle recovery $\gamma_c$, and combining both interventions on infected cattle dynamics $I_c(t)$}
\label{fig:jointcontrol}
\end{figure}

To assess the impact of combined control, we compared four scenarios: the baseline case with no intervention, a single intervention reducing $\beta_c$, a single intervention increasing $\gamma_c$, and a joint intervention combining both control measures. \myEdit{For the illustrative single-control and joint-control comparisons, we set $u_1=u_2=0.40$. Thus, the cattle-to-cattle transmission rate was reduced by $40\%$, the cattle recovery rate was increased by $40\%$, and the joint-control scenario applied both changes simultaneously. The resulting reproduction numbers were $4.423$ for the baseline, $2.700$ for the $\beta_c$-only intervention, $3.181$ for the
$\gamma_c$-only intervention, and $1.947$ for the joint intervention.} All other parameters were held fixed at their baseline values. The resulting infected cattle trajectories are shown in Fig.~\ref{fig:jointcontrol}.

Figure~\ref{fig:jointcontrol} shows that reducing $\beta_c$ alone lowers the epidemic peak relative to the baseline trajectory, confirming the important role of cattle--cattle transmission in sustaining outbreaks. Increasing $\gamma_c$ alone also reduces the infected cattle population by shortening the infectious period. However, the combined intervention produces the strongest reduction in $I_c(t)$, leading to a substantially lower epidemic peak and faster suppression of infection compared with either single intervention. This indicates that combining transmission reduction with enhanced recovery produces a greater reduction than either intervention applied alone.

The joint control response demonstrates that the model is more sensitive to simultaneous perturbations of transmission and recovery parameters than to isolated changes in either parameter. This is consistent with the structure of $\mathcal R_0$, in which $\beta_c$ appears in the numerator of the cattle transmission contribution, whereas $\gamma_c$ appears in the denominator through the infectious period. Hence, reducing $\beta_c$ and increasing $\gamma_c$ act in complementary directions to reduce the cattle contribution to $\mathcal R_0$.

\begin{figure}[ht!]
\centering
\includegraphics[width=0.5\linewidth]{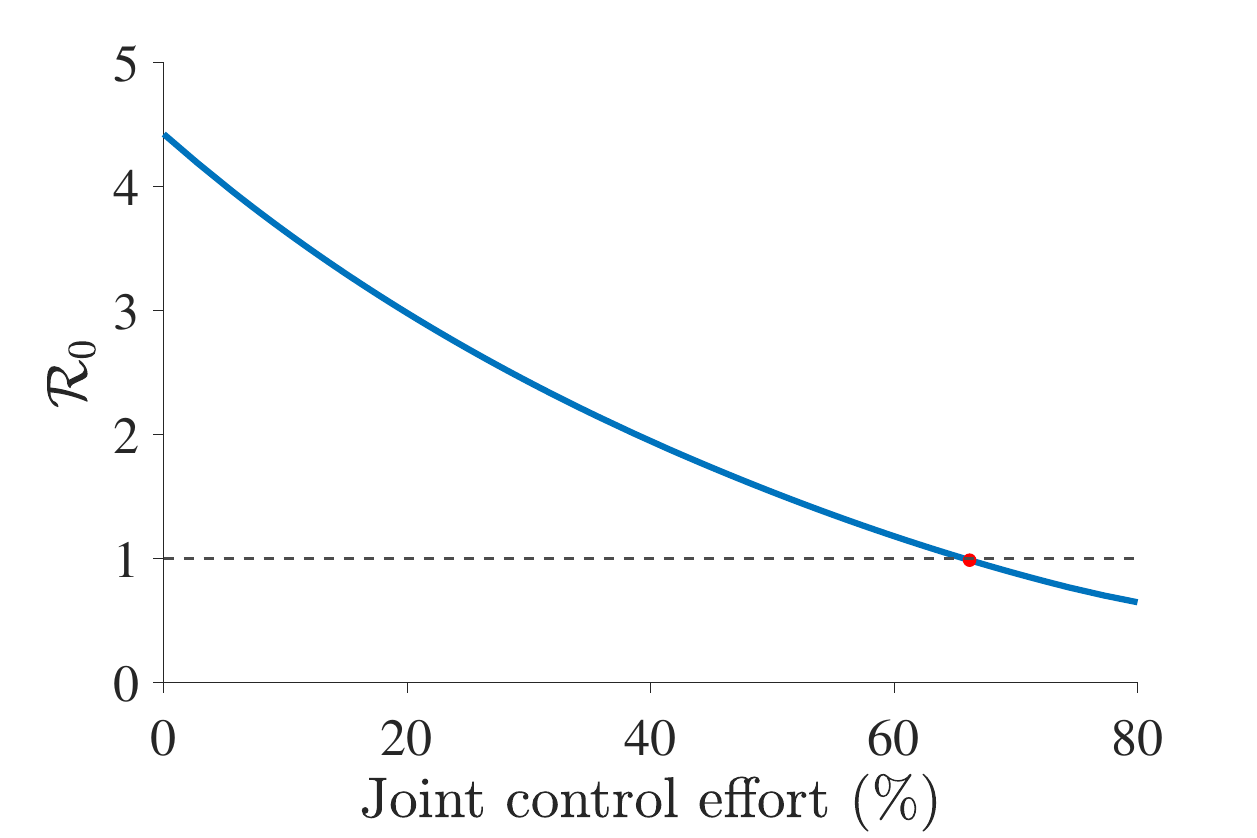}
\caption{Basic reproduction number $\mathcal R_0$ as a function of
joint-control effort. At control level $u$, the cattle transmission
and recovery parameters are modified as
$\beta_c\mapsto(1-u)\beta_c$ and
$\gamma_c\mapsto(1+u)\gamma_c$, respectively. The horizontal dashed
line indicates the invasion threshold $\mathcal R_0=1$.}
\label{fig:joint_control_R0}
\end{figure}

\myEdit{To complement the infected-cattle trajectories, we also evaluated how
the basic reproduction number changes with increasing joint-control
effort. Fig.~\ref{fig:joint_control_R0} shows that $\mathcal R_0$
decreases monotonically as $\beta_c$ is reduced and $\gamma_c$ is
increased simultaneously.}

\subsubsection{Marginal benefit and diminishing returns of joint control}

Building on the joint control strategy defined in the previous section, we further assessed the efficiency of coordinated intervention by varying the overall control effort and quantifying its effect on peak cattle infection. For each control level, the model was simulated and the percentage reduction in peak infected cattle, relative to the baseline scenario, was computed.

\begin{figure*}[ht!]
\centering
\begin{subfigure}[b]{0.48\linewidth}
    \centering
       \caption{}
    \includegraphics[width=\textwidth]{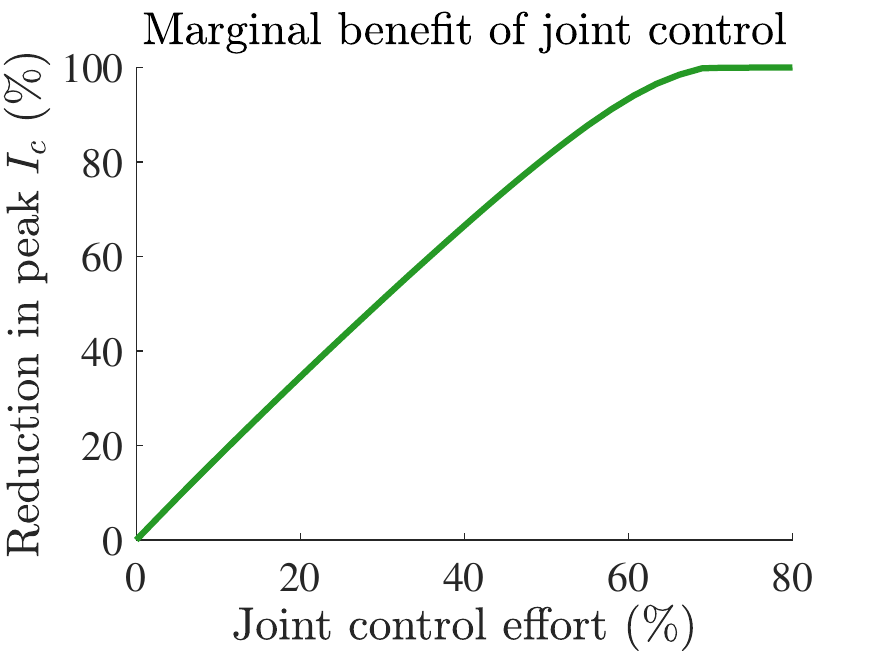}
    \label{fig:marginal_joint}
\end{subfigure}
\hfill
\begin{subfigure}[b]{0.48\linewidth}
    \centering
      \caption{}
    \includegraphics[width=\textwidth]{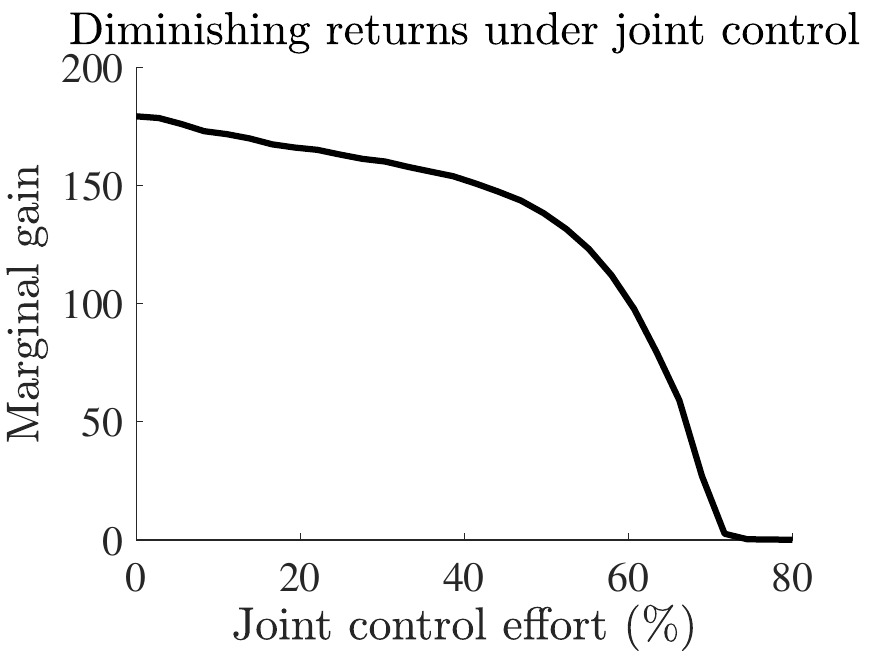}
     \label{fig:diminishing_joint}
\end{subfigure}
\caption{Cumulative and marginal benefits of the joint-control
strategy. (a) Cumulative percentage reduction in peak infected cattle
$I_c$ relative to the fixed baseline as joint-control effort
increases. (b) Marginal percentage-point reduction in peak $I_c$ per
additional one-percentage-point increase in joint-control effort. A
declining marginal-gain curve indicates diminishing returns.}
\label{fig:marginal_diminishing_joint}
\end{figure*}

Figure~\ref{fig:marginal_joint} shows that increasing joint control effort leads to a substantial reduction in the peak value of $I_c$. The response is nonlinear, with moderate control levels already producing large reductions in epidemic burden. This indicates that coordinated reductions in cattle transmission and increases in cattle recovery can substantially suppress outbreak severity without requiring maximal intervention intensity.

To examine whether additional control effort continues to provide proportional benefit, we also evaluated the marginal gain associated with increasing the joint control level. As shown in Fig.~\ref{fig:diminishing_joint}, the marginal gain decreases as control effort increases, indicating diminishing returns. Thus, while stronger intervention remains beneficial, the incremental improvement obtained from each additional increase in control effort becomes progressively smaller.

These results suggest that coordinated control strategies can achieve substantial epidemiological benefit at moderate implementation levels. From a practical perspective, this highlights the importance of balancing additional control effort against the expected marginal reduction in outbreak burden.

\section{Conclusion} \label{sec5}

This study developed and analysed a deterministic multi-host model for highly pathogenic avian influenza transmission at the cattle--wild bird--environment interface. The model incorporates direct transmission within cattle and wild birds, cross-species transmission from infected wild birds to cattle, and indirect transmission through a shared contaminated environmental reservoir. It was formulated as a population-level framework under homogeneous-mixing assumptions to investigate qualitative transmission mechanisms, threshold behaviour, and control-oriented scenarios.

Analytical results established positivity and boundedness of solutions, characterised the disease-free and endemic equilibria, and derived the basic reproduction number $\mathcal R_0$ using the next-generation matrix approach. \myEdit{The basic reproduction number was derived as the spectral radius of a reduced next-generation matrix whose entries represent within-cattle, within-bird, cross-species, and environmentally mediated transmission pathways. This formulation clarified how coupling among cattle, wild
birds, and the shared environment determines invasion potential.} Stability analysis confirmed the threshold behaviour of the system, with disease extinction associated with $\mathcal R_0<1$ and persistence possible when $\mathcal R_0>1$.

Numerical simulations supported the analytical findings and showed that long-term dynamics are primarily governed by the threshold properties of the model rather than by initial conditions. Global sensitivity analysis identified cattle recovery $\gamma_c$ and cattle--cattle transmission $\beta_c$ as the most influential parameters affecting $\mathcal R_0$, followed by environmental decay $\mu_{\mathrm{env}}$ and environmental transmission $\beta_{\mathrm{env}}$. These results indicate that direct cattle transmission and infectious-period duration form the dominant control axis, while environmental persistence acts as an important amplifying mechanism.

Surface plots and intervention scenarios further demonstrated that reducing cattle-to-cattle transmission produces the strongest reduction in epidemic burden, substantially lowering $\mathcal R_0$ and the peak number of infected cattle. Increasing cattle recovery or removal also reduced outbreak severity, although less strongly than reducing direct cattle transmission. In contrast, reducing environmental transmission alone produced only marginal changes in cattle infection trajectories under the baseline parameterisation, suggesting that environmental management is best viewed as a complementary control strategy rather than a standalone intervention.

\myEdit{The joint-control analysis showed that, under the assumed parameterisation, combining reductions in cattle transmission with increases in cattle recovery was more effective than applying either intervention alone. Marginal-benefit analyses further indicated that moderate levels of coordinated control could yield substantial reductions in outbreak burden, while additional intervention effort may produce diminishing returns. Within the assumptions and
parameter ranges explored by the model, these findings suggest that combining biosecurity and contact-reduction measures with rapid detection, isolation, or removal of infected cattle may provide greater epidemiological benefits than implementing either approach alone.}

The model has several limitations. \myEdit{The homogeneous mixing assumption
does not capture contact clustering within cattle herds or shared management
settings, such as barns and milking parlours. It also does not explicitly
represent spatial variation in cattle--wild-bird overlap or the seasonal and
migratory behaviour of wild birds. Consequently, the transmission parameters
should be interpreted as population-averaged effective rates, and the results
should not be viewed as farm-specific or spatially explicit predictions.}

\myEdit{The use of a single environmental compartment with a constant decay
rate is also a simplification. The parameter $\mu_{\mathrm{env}}$ represents
an effective average decay or removal rate and does not capture variation in
viral persistence across environmental substrates, locations, seasons, or
climatic conditions. Consequently, the environmental results should not be
interpreted as predictions for a specific substrate or field setting. Future
extensions could incorporate substrate-specific environmental compartments
and time-varying decay rates.}

\myEdit{The model further assumes that cattle in the exposed class $E_c$ are
not infectious and do not shed virus into the environment. Although the
infectious class $I_c$ may include preclinical or asymptomatic cattle, the
timing of viral shedding relative to infection and the onset of clinical signs
remains uncertain. If shedding begins during part of the stage represented by
$E_c$, the present formulation may underestimate early transmission and
environmental contamination. Future extensions could assign reduced
infectiousness and shedding rates to exposed cattle or divide the infection
process into non-infectious latent, preclinical infectious, and clinical
infectious stages.}

\myEdit{The model also assumes that recovered cattle and wild birds remain immune over
the time horizon considered. Because the duration of post-infection immunity to
HPAI H5N1 in these hosts is not well established, the model does not
represent the possible return of recovered animals to susceptibility. Waning
immunity would replenish the susceptible populations and could modify the
existence and stability of the endemic equilibrium. Under some parameter
regimes, this additional feedback may also produce transient or sustained
oscillations. Future extensions could incorporate immunity-loss transitions
from $R_c$ to $S_c$ and from $R_b$ to $S_b$ to investigate the effects of temporary immunity.}

\myEdit{The model does not account for external viral introduction.
Specifically, the recruitment term $\Lambda_b$ represents the entry of
susceptible wild birds only and excludes the arrival of infected migratory
birds or virus introduced from external environmental sources. Thus, the
extinction result applies only when no further infection is introduced from
outside the modelled system. Seasonal migration provides a potential mechanism
for repeated viral introductions and the initiation of new outbreaks following
local elimination. Future extensions could incorporate periodic migration,
infected-bird immigration, or a time-dependent external force of infection to
examine the effects of recurrent viral introduction.} The model also does not explicitly include cattle movement networks or
stochastic transmission processes, and the parameter distributions were used
for uncertainty exploration rather than data-driven inference.

\myEdit{Because the model was not calibrated to outbreak-specific
surveillance data, the quantitative results and biological
interpretations are conditional on the assumed model structure,
baseline parameterisation, and parameter ranges. Accordingly, the
intervention scenarios provide qualitative comparisons of the
relative effectiveness of the control measures within the theoretical
framework considered, rather than predictions for a specific outbreak
or host population. Nevertheless, the findings could inform future
policy discussions when considered alongside empirical evidence,
local epidemiological conditions, and practical implementation
constraints.}

Future work may extend the model by incorporating farm networks, cattle
movement, spatial structure, stochastic spillover, seasonality, empirical
surveillance data, and more detailed cattle infection stages. The intervention
scenarios developed here also provide a foundation for future optimal control
analysis, in which time-dependent controls could be used to investigate
cost-effective strategies for reducing cattle transmission, enhancing recovery
or removal, and limiting environmental contamination.

\section*{Code Availability}
The code used to generate the results and figures in this study is available at: 
\url{https://github.com/hamfat/HPAI-Cattle_Wildlife_Environment}

\section*{Competing/conflict of Interest declaration}
The authors declare that they have no known competing financial interests or personal relationships that could have appeared to influence the work reported in this paper.


\begin{thebibliography}{99}
 	
 	\bibitem{cdcAvianInfluenza} Centers for Disease Control and Prevention. \newblock \emph{Avian Influenza in Birds: Causes and How It Spreads}. \newblock 2024. \newblock Available at: \url{https://archive.cdc.gov/#/details?url=https://www.cdc.gov/bird-flu/virus-transmission/avian-in-birds.html}. Accessed 2025-03-04. \bibitem{LuSt15} Luczo, J.~M., Stambas, J., Durr, P.~A., Michalski, W.~P., and Bingham, J. \newblock Molecular pathogenesis of H5 highly pathogenic avian influenza: the role of the haemagglutinin cleavage site motif. \newblock \emph{Reviews in Medical Virology}, 25(6):406--430, 2015. \bibitem{SiRe25} Simancas-Racines, A., Reytor-González, C., Toral, M., and Simancas-Racines, D. \newblock H5N1 Avian Influenza: A Narrative Review of Scientific Advances and Global Policy Challenges. \newblock \emph{Viruses}, 17(7):927, 2025. \bibitem{cdcAvianInfluenza2} Centers for Disease Control and Prevention. \newblock \emph{Bird Flu: Causes and How It Spreads}. \newblock 2022. \newblock Available at: \url{https://www.cdc.gov/bird-flu/virus-transmission/index.html}. Accessed 2025-03-05. \bibitem{mpi1} Ministry for Primary Industries. \newblock \emph{Avian Influenza (HPAI) and Dairy Cows}. \newblock NZ Government, 2025. \newblock Available at: \url{https://www.mpi.govt.nz/biosecurity/pest-and-disease-threats-to-new-zealand/animal-disease-threats-to-new-zealand/high-pathogenicity-avian-influenza/avian-influenza-dairy-cattle-and-other-livestock/}. Accessed 2025-03-05. \bibitem{KaHo08} Kalthoff, D., Hoffmann, B., Harder, T., Durban, M., and Beer, M. \newblock Experimental infection of cattle with highly pathogenic avian influenza virus (H5N1). \newblock \emph{Emerging Infectious Diseases}, 14(7):1132, 2008. \bibitem{ChRu23} Charostad, J., Rukerd, M.~R.~Z., Mahmoudvand, S., Bashash, D., Hashemi, S.~M.~A., Nakhaie, M., and Zandi, K. \newblock A comprehensive review of highly pathogenic avian influenza (HPAI) H5N1: an imminent threat at doorstep. \newblock \emph{Travel Medicine and Infectious Disease}, 55:102638, 2023. \bibitem{LiDu08} Liu, R., Duvvuri, V.~R.~S.~K., and Wu, J. \newblock Spread Pattern Formation of H5N1-Avian Influenza and its Implications for Control Strategies. \newblock \emph{Mathematical Modelling of Natural Phenomena}, 3(7):161--179, 2008. \bibitem{HaMc25} Hassman, R.~L., McCabe, I.~M.~H., Smith, K.~M., and Allen, L.~J.~S. \newblock Stochastic Models of Zoonotic Avian Influenza with Multiple Hosts, Environmental Transmission, and Migration in the Natural Reservoir. \newblock \emph{Bulletin of Mathematical Biology}, 87(1):1--43, 2025. \bibitem{Nevada} Central Nevada Health District. \newblock \emph{The Central Nevada Health District is Actively Monitoring for Spread of H5N1 in Northern Nevada}. \newblock 2025. \newblock Available at: \url{https://www.centralnevadahd.org/press-release/}. Accessed 2025-03-05. \bibitem{cdcAvianInfluenza3} Centers for Disease Control and Prevention. \newblock \emph{CDC A(H5N1) Bird Flu Response Update February 26, 2025}. \newblock 2025. \newblock Available at: \url{https://www.cdc.gov/bird-flu/spotlights/h5n1-response-02262025.html}. Accessed 2025-03-05. \bibitem{IwTa07} Iwami, S., Takeuchi, Y., and Liu, X. \newblock Avian-human influenza epidemic model. \newblock \emph{Mathematical Biosciences}, 207(1):1--25, 2007. \bibitem{GaHa11} Garner, M.~G. and Hamilton, S.~A. \newblock Principles of epidemiological modelling. \newblock \emph{Revue Scientifique et Technique}, 30(2):407, 2011. \bibitem{WaFe08} Wang, H., Feng, Z., Shu, Y., Yu, H., Zhou, L., Zu, R., Huai, Y., Dong, J., Bao, C., Wen, L., et al. \newblock Probable limited person-to-person transmission of highly pathogenic avian influenza A (H5N1) virus in China. \newblock \emph{The Lancet}, 371(9622):1427--1434, 2008. \bibitem{BoGo11} Bourouiba, L., Gourley, S.~A., Liu, R., and Wu, J. \newblock The interaction of migratory birds and domestic poultry and its role in sustaining avian influenza. \newblock \emph{SIAM Journal on Applied Mathematics}, 71(2):487--516, 2011. \bibitem{ShMo18} Sharma, S., Mondal, A., Pal, A.~K., and Samanta, G.~P. \newblock Stability analysis and optimal control of avian influenza virus A with time delays. \newblock \emph{International Journal of Dynamics and Control}, 6:1351--1366, 2018. \bibitem{PaBu13} Pandit, P.~S., Bunn, D.~A., Pande, S.~A., and Aly, S.~S. \newblock Modeling highly pathogenic avian influenza transmission in wild birds and poultry in West Bengal, India. \newblock \emph{Scientific Reports}, 3(1):2175, 2013. \bibitem{DuEl24} Duarte, P.~M., El-Nakeep, S., Shayestegan, F., Tazerji, S.~S., Malik, Y.~S., Roncada, P., Tilocca, B., Gharieb, R., Hogan, U., Ahmadi, H., et al. \newblock Addressing the recent transmission of H5N1 to new animal species and humans, warning of the risks and its relevance in One Health. \newblock \emph{German Journal of Microbiology}, 4(2):39--53, 2024.
 	
 	\bibitem{YeLi20} Ye, X., Lin, S., and Xu, C. \newblock Dynamical analysis of a fractional-order avian-human influenza epidemic model with logistic growth for avian population. \newblock \emph{Journal of Algorithms and Computational Technology}, 14:1748302620966704, 2020. \bibitem{farman2024global} Farman, M., Alfiniyah, C., and Saqib, M. \newblock Global Stability with Lyapunov Function and Dynamics of SEIR-Modified Lassa Fever Model in Sight Power Law Kernel. \newblock \emph{Complexity}, 2024(1):3562684, 2024. \bibitem{mayengo2023volterra} Mayengo, M.~M. \newblock The Volterra--Lyapunov matrix theory for global stability analysis of alcohol-related health risks model. \newblock \emph{Results in Physics}, 44:106149, 2023. \bibitem{MATLAB2025} The MathWorks Inc. \newblock \emph{MATLAB Version R2025a}. \newblock The MathWorks Inc., Natick, Massachusetts, USA, 2025. \bibitem{BUTT2024S13} Butt, S.~L., Nooruzzaman, M., Covaleda, L.~M., and Diel, D.~G. \newblock Hot topic: Influenza A H5N1 virus exhibits a broad host range, including dairy cows. \newblock \emph{JDS Communications}, 5(Suppl.~1):S13--S19, 2024. \bibitem{dunning2025spread} Dunning, J., Firth, J.~A., and Ward, A.~I. \newblock The spread of highly pathogenic avian influenza virus is a social network problem. \newblock \emph{PLoS Pathogens}, 21(7):e1013233, 2025. \bibitem{Holt06} Holt, J., Davis, S., and Leirs, H. \newblock A model of leptospirosis infection in an African rodent to determine risk to humans: Seasonal fluctuations and the impact of rodent control. \newblock \emph{Acta Tropica}, 99(2--3):218--225, 2006. \bibitem{Fatoyinbo2025} Fatoyinbo, H.~O., Tiwari, P., Ip, R.~H.~L., and Miranda, V. \newblock Multivariable analysis of highly pathogenic H5N1 and H5Nx avian influenza in wild birds and poultry in Asian subregions. \newblock \emph{Comparative Immunology, Microbiology and Infectious Diseases}, 2025. \bibitem{stanislawek2024} Stanislawek, W.~L., Tana, T., Rawdon, T.~G., Cork, S.~C., Chen, K., Fatoyinbo, H., Cogger, N., Webby, R.~J., Webster, R.~G., Joyce, M., Tuboltsev, M.~A., Orr, D., Ohneiser, S., Watts, J., Riegen, A.~C., McDougall, M., Klee, D., and O'Keefe, J.~S. \newblock Avian influenza viruses in New Zealand wild birds, with an emphasis on subtypes H5 and H7: Their distinctive epidemiology and genomic properties. \newblock \emph{PLOS ONE}, 19(6):e0303756, 2024. \bibitem{Esaki25} Esaki, M., Okuya, K., Tokorozaki, K., Haraguchi, Y., Ito, J., and Ozawa, M. \newblock Surveillance of avian influenza viruses in the Izumi plain reveals the role of wild ducks in the introduction of H5N1 HPAIVs during the 2023/24 winter season. \newblock \emph{Comparative Immunology, Microbiology and Infectious Diseases}, 123:102389, 2025. \bibitem{Fereidouni23} Fereidouni, S., Starick, E., Karamendin, K., Di Genova, C., Scott, S.~D., Khan, Y., Harder, T., and Kydyrmanov, A. \newblock Genetic characterization of a new candidate hemagglutinin subtype of influenza A viruses. \newblock \emph{Emerging Microbes \& Infections}, 12(2):2225645, 2023. \bibitem{sangrat24} Sangrat, W., Thanapongtharm, W., Kasemsuwan, S., Boonyawiwat, V., Sajapitak, S., and Poolkhet, C. \newblock Geospatial and Temporal Analysis of Avian Influenza Risk in Thailand: A GIS-Based Multi-Criteria Decision Analysis Approach for Enhanced Surveillance and Control. \newblock \emph{Transboundary and Emerging Diseases}, 2024(1):6474182, 2024. \bibitem{vreman23} Vreman, S., Kik, M., Germeraad, E., Heutink, R., Harders, F., Spierenburg, M., Engelsma, M., Rijks, J., van den Brand, J., and Beerens, N. \newblock Zoonotic Mutation of Highly Pathogenic Avian Influenza H5N1 Virus Identified in the Brain of Multiple Wild Carnivore Species. \newblock \emph{Pathogens}, 12(2):168, 2023. \bibitem{Bellotti2024} Bellotti, B.~R., DeWitt, M.~E., Wenner, J.~J., Lombard, J.~E., McCluskey, B.~J., and Kortessis, N. \newblock Challenges and lessons learned from preliminary modeling of within-herd transmission of highly pathogenic avian influenza H5N1 in dairy cattle. \newblock \emph{bioRxiv}, 2024. \bibitem{Rawson2025} Rawson, T., Morgenstern, C., Knock, E.~S., Hicks, J., Pham, A., Morel, G., Murillo, C.~A., Sanderson, M.~W., Forchini, G., FitzJohn, R., Hauck, K., and Ferguson, N. \newblock A mathematical model of H5N1 influenza transmission in US dairy cattle. \newblock \emph{Nature Communications}, 16(1):4308, 2025. \bibitem{gumel2009} Gumel, A.~B. \newblock Global dynamics of a two-strain avian influenza model. \newblock \emph{International Journal of Computer Mathematics}, 86(1):85--108, 2009. \bibitem{Regassa2024} Regassa, A.~G. and Obsu, L.~L. \newblock The role of asymptomatic cattle for leptospirosis dynamics in a herd with imperfect vaccination. \newblock \emph{Scientific Reports}, 14:23775, 2024.
 	
 	 \bibitem{ABIDEMI2022} Abidemi, A., Ackora-Prah, J., Fatoyinbo, H.~O., and Asamoah, J.~K.~K. \newblock Lyapunov stability analysis and optimization measures for a dengue disease transmission model. \newblock \emph{Physica A}, 602:127646, 2022. 
 	
 	\bibitem{BlTr21} Blagodatski, A., Trutneva, K., Glazova, O., Mityaeva, O., Shevkova, L., Kegeles, E., Onyanov, N., Fede, K., Maznina, A., Khavina, E., Yeo, S., Park, H., and Volchkov, P. \newblock Avian influenza in wild birds and poultry: dissemination pathways, monitoring methods, and virus ecology. \newblock \emph{Pathogens}, 10(5):630, 2021.
 	
 	 \bibitem{GiJa24} Giacinti, J.~A., Jarvis-Cross, M., Lewis, H., Provencher, J.~F., Berhane, Y., Kuchinski, K., Jardine, C.~M., Signore, A., Mansour, S.~C., Sadler, D.~E., et al. \newblock Transmission dynamics of highly pathogenic avian influenza virus at the wildlife-poultry-environmental interface: a case study. \newblock \emph{One Health}, 19:100932, 2024.
 	
 	 \bibitem{SoHe25} Song, W., He, F., Deng, Z., Hu, M., Fang, K., Cheng, W., Wu, J., Wang, X., Fan, G., Kong, L., et al. \newblock Avian influenza virus dynamics in poultry and the environment: an eight-year longitudinal study in the southwestern Poyang Lake region of China. \newblock \emph{Infectious Disease Modelling}, 2025.
 	
 	\bibitem{AyKh24} Ayuti, S.~R., Khairullah, A.~R., Lamid, M., Al-Arif, M.~A., Warsito, S.~H., Silaen, O.~S.~M., Moses, I.~B., Hermawan, I.~P., Yanestria, S.~M., Delima, M., et al. \newblock Avian influenza in birds: Insights from a comprehensive review. \newblock \emph{Veterinary World}, 17(11):2544, 2024. \bibitem{GrLu24} Graziosi, G., Lupini, C., Catelli, E., and Carnaccini, S. \newblock Highly pathogenic avian influenza (HPAI) H5 clade 2.3.4.4b virus infection in birds and mammals. \newblock \emph{Animals}, 14(9):1372, 2024. \bibitem{MaLe25} Martí-Garcia, B., Lean, F.~Z.~X., Núñez, A., and Majó, N. \newblock Natural infections of highly pathogenic avian influenza virus H5N1 in wild birds between 2020 and 2023 in the UK: a retrospective study with focus on microscopic lesions, viral distribution and neurotropism. \newblock \emph{Veterinary Research}, 56(1):218, 2025. \bibitem{BoVr23} Bordes, L., Vreman, S., Heutink, R., Roose, M., Venema, S., Pritz-Verschuren, S.~B.~E., Rijks, J.~M., Gonzales, J.~L., Germeraad, E.~A., Engelsma, M., et al. \newblock Highly pathogenic avian influenza H5N1 virus infections in wild red foxes (Vulpes vulpes) show neurotropism and adaptive virus mutations. \newblock \emph{Microbiology Spectrum}, 11(1):e02867--22, 2023. \bibitem{PlGa24} Plaza, P.~I., Gamarra-Toledo, V., Euguí, J.~R., and Lambertucci, S.~A. \newblock Recent changes in patterns of mammal infection with highly pathogenic avian influenza A (H5N1) virus worldwide. \newblock \emph{Emerging Infectious Diseases}, 30(3):444, 2024. \bibitem{NiCh24} Niedringhaus, K.~D., Chan, T.~C., McDowell, A., Maxwell, L., Stevens, M., Potts, L., Miller, E., Anis, E., Why, K.~V., Keller, T., et al. \newblock Surveillance of Highly Pathogenic Avian Influenza Virus in Wild Canids from Pennsylvania, USA. \newblock \emph{Animals}, 14(24):3700, 2024. \bibitem{CaFr24} Caserta, L.~C., Frye, E.~A., Butt, S.~L., Laverack, M., Nooruzzaman, M., Covaleda, L.~M., Thompson, A.~C., Koscielny, M.~P., Cronk, B., Johnson, A., et al. \newblock Spillover of highly pathogenic avian influenza H5N1 virus to dairy cattle. \newblock \emph{Nature}, 634(8034):669--676, 2024. \bibitem{SaBo25} Sanchez-Rojas, I.~C., Bonilla-Aldana, D.~K., Solarte-Jimenez, C.~L., Bonilla-Aldana, J.~L., Acosta-España, J.~D., and Rodriguez-Morales, A.~J. \newblock Highly Pathogenic Avian Influenza (H5N1) Clade 2.3.4.4b in Cattle: A Rising One Health Concern. \newblock \emph{Animals}, 15(13):1963, 2025. \bibitem{RoPi24} Rodriguez, Z., Picasso-Risso, C., O'Connor, A., and Ruegg, P.~L. \newblock Hot topic: Epidemiological and clinical aspects of highly pathogenic avian influenza H5N1 in dairy cattle. \newblock \emph{JDS Communications}, 5:S8--S12, 2024. \bibitem{BaCo25} Banyard, A.~C., Coombes, H., Terrey, J., McGinn, N., Seekings, J., Clifton, B., Mollett, B.~C., Di Genova, C., Sainz-Dominguez, P., Worsley, L., et al. \newblock Detection of clade 2.3.4.4b H5N1 high pathogenicity avian influenza virus in a sheep in Great Britain, 2025. \newblock \emph{Emerging Microbes \& Infections}, 14(1):2547730, 2025. \bibitem{KiJe24} Kim, J., Jeong, S., Kim, D., Lee, D., Lee, D., Kim, D., and Kwon, J. \newblock Genomic epidemiology of highly pathogenic avian influenza A (H5N1) virus in wild birds in South Korea during 2021--2022: Changes in viral epidemic patterns. \newblock \emph{Virus Evolution}, 10(1):veae014, 2024. \bibitem{DuIs22} Dutta, P., Islam, A., Sayeed, M.~A., Rahman, M.~A., Abdullah, M.~S., Saha, O., Rahman, M.~Z., Klaassen, M., Hoque, M.~A., and Hassan, M.~M. \newblock Epidemiology and molecular characterization of avian influenza virus in backyard poultry of Chattogram, Bangladesh. \newblock \emph{Infection, Genetics and Evolution}, 105:105377, 2022. \bibitem{MoAb24} Moatasim, Y., Aboulhoda, B.~E., Gomaa, M., El Taweel, A., Kutkat, O., Kamel, M.~N., El Sayes, M., GabAllah, M., Elkhrsawy, A., AbdAllah, H., et al. \newblock Genetic and pathogenic potential of highly pathogenic avian influenza H5N8 viruses from live bird markets in Egypt in avian and mammalian models. \newblock \emph{PLoS ONE}, 19(10):e0312134, 2024. \bibitem{LaBa23} Lambert, S., Bauzile, B., Mugnier, A., Durand, B., Vergne, T., and Paul, M.~C. \newblock A systematic review of mechanistic models used to study avian influenza virus transmission and control. \newblock \emph{Veterinary Research}, 54(1):96, 2023. \bibitem{Mostafa2025HPAI} Mostafa, A., Nogales, A., and Martinez-Sobrido, L. \newblock Highly pathogenic avian influenza H5N1 in the United States: recent incursions and spillover to cattle. \newblock \emph{npj Viruses}, 3:54, 2025. \newblock doi: 10.1038/s44298-025-00138-5. \bibitem{HPAI} U.S. Department of Agriculture. \newblock \emph{HPAI in Livestock | Animal and Plant Health Inspection Service}. \newblock Available at: \url{https://www.aphis.usda.gov/livestock-poultry-disease/avian/avian-influenza/hpai-livestock}. Accessed 2025-03-05. \bibitem{WOAH_AvianInfluenza} World Organisation for Animal Health (WOAH). \newblock \emph{Avian Influenza}. \newblock 2024. \newblock Available at: \url{https://www.woah.org/en/disease/avian-influenza/}. Accessed 2025-11-28. \bibitem{Nguyen2020H5Vietnam} Nguyen, L.~T., Stevenson, M.~A., Firestone, S.~M., Sims, L.~D., Chu, D.~H., Nguyen, L.~V., Nguyen, T.~N., Le, K.~T., Isoda, N., Matsuno, K., Okamatsu, M., Kida, H., and Sakoda, Y. \newblock Spatiotemporal and risk analysis of H5 highly pathogenic avian influenza in Vietnam, 2014--2017. \newblock \emph{Preventive Veterinary Medicine}, 178:104678, 2020. \newblock doi: 10.1016/j.prevetmed.2019.04.007.
 	
 	\bibitem{Abidemi2022Wolbachia} Abidemi, A., Fatoyinbo, H.~O., and Asamoah, J.~K.~K. \newblock Analysis of Dengue Fever Transmission Dynamics with Multiple Controls: A Mathematical Approach. \newblock In \emph{Proceedings of the 2020 International Conference on Decision Aid Sciences and Application (DASA)}, pages 971--978, 2020. \newblock doi: 10.1109/DASA51403.2020.9317064. \bibitem{AbidemiCovid} Abidemi, A. and Fatoyinbo, H.~O. \newblock Mathematical Analysis of Optimal Cost-Effective Control of COVID-19: A Case Study. \newblock In \emph{Proceedings of the 2021 International Conference on Decision Aid Sciences and Application (DASA)}, pages 95--102, 2021. \newblock doi: 10.1109/DASA53625.2021.9682382. \bibitem{Deng2024HIVModel} Deng, Q., Guo, T., Qiu, Z., et al. \newblock A mathematical model for HIV dynamics with multiple infections: implications for immune escape. \newblock \emph{Journal of Mathematical Biology}, 89(6), 2024. \newblock doi: 10.1007/s00285-024-02104-w. \bibitem{Rajnarayanan2025MalariaModel} Rajnarayanan, A., Kumar, M., and Tridane, A. \newblock Analysis of a mathematical model for malaria using data-driven approach. \newblock \emph{Scientific Reports}, 15:27272, 2025. \newblock doi: 10.1038/s41598-025-12078-4. \bibitem{USDALivestock} Animal and Plant Health Inspection Service, U.S. Department of Agriculture. \newblock \emph{HPAI Confirmed Cases in Livestock}. \newblock Available at: \url{https://www.aphis.usda.gov/livestock-poultry-disease/avian/avian-influenza/hpai-detections/hpai-confirmed-cases-livestock}. Accessed 2026-01-16. \bibitem{USDAWild_Birds} Animal and Plant Health Inspection Service, U.S. Department of Agriculture. \newblock \emph{Detections of Highly Pathogenic Avian Influenza in Wild Birds}. \newblock Available at: \url{https://www.aphis.usda.gov/livestock-poultry-disease/avian/avian-influenza/hpai-detections/wild-birds}. Accessed 2026-01-16. \bibitem{Rejuetal24} John, R.~S., Fatoyinbo, H.~O., and Hayman, D.~T.~S. \newblock Modelling Lassa virus dynamics in West African Mastomys natalensis and the impact of human activities. \newblock \emph{Journal of the Royal Society Interface}, 20240106, 2024. \newblock doi: 10.1098/rsif.2024.0106. \bibitem{CHANG2026} Chang, Y., Gonzales, J.~L., Rattenborg, E., de Jong, M.~C.~M., and Conrady, B. \newblock Modeling the spillover risk of highly pathogenic avian influenza from wild birds to cattle in Denmark: A data-driven risk assessment framework. \newblock \emph{Preventive Veterinary Medicine}, 251:106844, 2026. \newblock doi: 10.1016/j.prevetmed.2026.106844. \bibitem{PenaMosca2025} Peña-Mosca, F., Frye, E.~A., MacLachlan, M.~J., Rebelo, A.~R., de Oliveira, P.~S.~B., Nooruzzaman, M., Koscielny, M.~P., Zurakowski, M., Lieberman, Z.~R., Leone, W.~M., Elvinger, F., Nydam, D.~V., and Diel, D.~G. \newblock The impact of highly pathogenic avian influenza H5N1 virus infection on dairy cows. \newblock \emph{Nature Communications}, 16:6520, 2025. \newblock doi: 10.1038/s41467-025-61553-z. \bibitem{AVMA2025} American Veterinary Medical Association. \newblock \emph{Avian Influenza Virus Type A (H5N1) in U.S. Dairy Cattle}. \newblock 2025. \newblock Available at: \url{https://www.avma.org/resources-tools/animal-health-and-welfare/animal-health/avian-influenza/avian-influenza-virus-type-h5n1-us-dairy-cattle}. Accessed 2026-05-01. \bibitem{Martin2018} Martin, G., Becker, D.~J., and Plowright, R.~K. \newblock Environmental persistence of influenza H5N1 is driven by temperature and salinity: Insights from a Bayesian meta-analysis. \newblock \emph{Frontiers in Ecology and Evolution}, 6:131, 2018. \newblock doi: 10.3389/fevo.2018.00131. \bibitem{DeVries2020} De Vries, A. \newblock Symposium review: Why revisit dairy cattle productive lifespan? \newblock \emph{Journal of Dairy Science}, 103(4):3838--3845, 2020. \newblock doi: 10.3168/jds.2019-17361. \bibitem{DrWa02} Van den Driessche, P. and Watmough, J. \newblock Reproduction numbers and sub-threshold endemic equilibria for compartmental models of disease transmission. \newblock \emph{Mathematical Biosciences}, 180(1--2):29--48, 2002. \bibitem{Ra11} Rantzer, A. \newblock Distributed control of positive systems. \newblock In \emph{Proceedings of the 50th IEEE Conference on Decision and Control and European Control Conference}, pages 6608--6611, 2011. \newblock IEEE.
 	
 	\end{thebibliography}

\appendix
\label{sec:app}

\section{Proof of Theorem~\ref{thm1}}

\begin{proof}
The vector field of the model is locally Lipschitz, so a unique local solution exists for each initial condition. It is sufficient to verify that the vector field is non-negative on each boundary face of the non-negative orthant.

For the cattle compartments, when the corresponding state variable is zero, we have
\[
\left.\frac{dS_c}{dt}\right|_{S_c=0}=\Lambda_c\geq 0,
\qquad
\left.\frac{dE_c}{dt}\right|_{E_c=0}
=\lambda_cS_c\geq 0,
\]
\[
\left.\frac{dI_c}{dt}\right|_{I_c=0}
=\sigma_cE_c\geq 0,
\qquad
\left.\frac{dR_c}{dt}\right|_{R_c=0}
=\gamma_cI_c\geq 0.
\]
Similarly, for the wild-bird compartments,
\[
\left.\frac{dS_b}{dt}\right|_{S_b=0}=\Lambda_b\geq 0,
\qquad
\left.\frac{dI_b}{dt}\right|_{I_b=0}
=\lambda_bS_b\geq 0,
\]
\[
\left.\frac{dR_b}{dt}\right|_{R_b=0}
=\gamma_bI_b\geq 0.
\]
Finally,
\[
\left.\frac{dB}{dt}\right|_{B=0}
=\theta_cI_c+\theta_bI_b\geq 0.
\]
Hence no component can cross from the non-negative orthant into the negative region. Therefore all state variables remain non-negative for all $t\geq 0$.

For completeness, the susceptible cattle equation may also be written in the integrating-factor form
\[
\frac{dS_c}{dt}
+\bigl(\lambda_c+\mu_{dc}\bigr)S_c
=\Lambda_c,
\]
which yields
\begin{align*}
S_c(t)
={}&\exp\!\left[-\int_0^t
\bigl(\lambda_c(z)+\mu_{dc}\bigr)\,dz\right]S_c(0)
+\int_0^t \Lambda_c
\exp\!\left[-\int_s^t
\bigl(\lambda_c(z)+\mu_{dc}\bigr)\,dz\right]ds
\\&\geq 0.
\end{align*}
The remaining compartments admit analogous computations.
\end{proof}

\section{Proof of Theorem~\ref{thm:1}}

\begin{proof}
Using the equations of the full model \eqref{eqn:fullmodel}, the total cattle population satisfies
\begin{align}
\frac{dN_c}{dt}
&=\frac{dS_c}{dt}+\frac{dE_c}{dt}+\frac{dI_c}{dt}+\frac{dR_c}{dt}\notag\\
&=\Lambda_c-\mu_{dc}N_c-\mu_c(E_c+I_c).
\label{eq:Nc}
\end{align}
Since $E_c(t),I_c(t)\geq 0$,
\[
\frac{dN_c}{dt}\leq \Lambda_c-\mu_{dc}N_c.
\]
Also, because $E_c+I_c\leq N_c$,
\[
\frac{dN_c}{dt}
\geq \Lambda_c-(\mu_c+\mu_{dc})N_c.
\]
Therefore, we have the two-sided differential inequality
\[
\Lambda_c-(\mu_c+\mu_{dc})N_c(t)
\leq N_c'(t)
\leq \Lambda_c-\mu_{dc}N_c(t).
\]
By comparison, the solution $N(t)$ satisfies
\begin{align*}
&\frac{\Lambda_c}{\mu_c+\mu_{dc}}
+e^{-(\mu_c+\mu_{dc})t}
\left(N_c(0)-\frac{\Lambda_c}{\mu_c+\mu_{dc}}\right)
\\
&\hspace{3cm}\leq N_c(t)
\leq
\frac{\Lambda_c}{\mu_{dc}}
+e^{-\mu_{dc}t}
\left(N_c(0)-\frac{\Lambda_c}{\mu_{dc}}\right).
\end{align*}
Consequently,
\[
\frac{\Lambda_c}{\mu_c+\mu_{dc}}
\leq \liminf_{t\to\infty}N_c(t)
\leq \limsup_{t\to\infty}N_c(t)
\leq \frac{\Lambda_c}{\mu_{dc}}.
\]
In particular,
\[
N_c(t)\leq M_c:=\max\left\{N_c(0),\frac{\Lambda_c}{\mu_{dc}}\right\}
\qquad \text{for all }t\geq 0.
\]

Similarly, for the total wild-bird population $N_b=S_b+I_b+R_b$,
\begin{align}
\frac{dN_b}{dt}
&=\Lambda_b-\mu_{db}N_b-\mu_bI_b.
\label{eq:Nb}
\end{align}
Hence
\[
\Lambda_b-(\mu_b+\mu_{db})N_b(t)
\leq N_b'(t)
\leq \Lambda_b-\mu_{db}N_b(t),
\]
which gives
\begin{align*}
&\frac{\Lambda_b}{\mu_b+\mu_{db}}
+e^{-(\mu_b+\mu_{db})t}
\left(N_b(0)-\frac{\Lambda_b}{\mu_b+\mu_{db}}\right)
\\
&\hspace{3cm}\leq N_b(t)
\leq
\frac{\Lambda_b}{\mu_{db}}
+e^{-\mu_{db}t}
\left(N_b(0)-\frac{\Lambda_b}{\mu_{db}}\right).
\end{align*}
Thus,
\[
\frac{\Lambda_b}{\mu_b+\mu_{db}}
\leq \liminf_{t\to\infty}N_b(t)
\leq \limsup_{t\to\infty}N_b(t)
\leq \frac{\Lambda_b}{\mu_{db}},
\]
and
\[
N_b(t)\leq M_b:=\max\left\{N_b(0),\frac{\Lambda_b}{\mu_{db}}\right\}
\qquad \text{for all }t\geq 0.
\]

Next, consider the environmental compartment. Since $I_c\leq N_c$ and $I_b\leq N_b$,
\begin{align*}
\frac{dB}{dt}
&=\theta_cI_c(t)+\theta_bI_b(t)-\mu_{\mathrm{env}}B(t)\\
&\leq \theta_cN_c(t)+\theta_bN_b(t)-\mu_{\mathrm{env}}B(t)\\
&\leq \theta_cM_c+\theta_bM_b-\mu_{\mathrm{env}}B(t).
\end{align*}
Therefore,
\[
B(t)\leq
\frac{\theta_cM_c+\theta_bM_b}{\mu_{\mathrm{env}}}
+
\left[
B(0)-\frac{\theta_cM_c+\theta_bM_b}{\mu_{\mathrm{env}}}
\right]e^{-\mu_{\mathrm{env}}t}.
\]
Hence $B(t)$ is bounded for all $t\geq 0$. It follows that all state variables of the full model, and therefore all variables of the reduced system \eqref{eqn:reduced}, are bounded.
\end{proof}

\section{Global stability analysis of equilibria}
\label{sec:GlobStab}

\subsection{Global stability of the disease-free equilibrium}
\myEdit{
The disease-free equilibrium of system~\eqref{eqn:reduced} is
\[
\psi_{\text{DFE}}
=
\left(
S_c^*,0,0,S_b^*,0,0
\right),
\]
where
\[
S_c^*=\frac{\Lambda_c}{\mu_{dc}},
\qquad
S_b^*=\frac{\Lambda_b}{\mu_{db}}.
\]

The positively invariant biologically feasible region
$\mathcal E$ satisfies
\[
0\leq S_c(t)\leq S_c^*,
\qquad
0\leq S_b(t)\leq S_b^*.
\]

Let
\[
X=
\begin{pmatrix}
E_c&I_c&I_b&B
\end{pmatrix}^{\mathsf T}
\]
denote the vector of infected-state variables. At a general state of
the system, define the new-infection matrix
\[
\widetilde{\mathcal F}(S_c,S_b)
=
\begin{pmatrix}
0 & \beta_cS_c & \beta_{bc}S_c
  & \beta_{\mathrm{env}}S_c\\
0 & 0 & 0 & 0\\
0 & 0 & \beta_bS_b
  & \beta_{\mathrm{env}}S_b\\
0 & 0 & 0 & 0
\end{pmatrix}.
\]
The infected subsystem can then be written as
\[
\dot X
=
\left[
\widetilde{\mathcal F}(S_c,S_b)-\mathcal V
\right]X,
\]
where
\[
\mathcal V
=
\begin{pmatrix}
k_E & 0 & 0 & 0\\
-\sigma_c & k_c & 0 & 0\\
0 & 0 & k_b & 0\\
0 & -\theta_c & -\theta_b & \mu_{\mathrm{env}}
\end{pmatrix},
\]
with
\[
k_E=\sigma_c+\mu_c+\mu_{dc},
\qquad
k_c=\gamma_c+\mu_c+\mu_{dc},
\qquad
k_b=\gamma_b+\mu_b+\mu_{db}.
\]

Let
\[
\mathcal F
=
\widetilde{\mathcal F}(S_c^*,S_b^*)
\]
denote the new-infection matrix evaluated at the disease-free
equilibrium. Since
\[
S_c(t)\leq S_c^*
\qquad\text{and}\qquad
S_b(t)\leq S_b^*,
\]
we have the component-wise inequality
\[
\widetilde{\mathcal F}(S_c,S_b)
\leq
\mathcal F.
\]
Because $X\geq0$, it follows that
\[
\dot X
\leq
(\mathcal F-\mathcal V)X
\]
component-wise.

Recall that the basic reproduction number is defined by
\[
\mathcal R_0
=
\rho(\mathcal F\mathcal V^{-1}),
\]
where $\rho(\cdot)$ denotes the spectral radius. The matrix
$\mathcal V$ is a nonsingular $M$-matrix, and the standard
next-generation matrix result implies that
\[
\mathcal R_0<1
\quad\Longleftrightarrow\quad
s(\mathcal F-\mathcal V)<0,
\]
where $s(\cdot)$ denotes the spectral bound, namely the maximum of
the real parts of the eigenvalues of a matrix~\cite{DrWa02}.
Consequently,
\[
s(\mathcal F-\mathcal V)<0
\]
implies that all eigenvalues of $\mathcal F-\mathcal V$ have negative
real parts. Therefore,
\[
\mathcal A=\mathcal F-\mathcal V
\]
is a Hurwitz matrix. Moreover, $\mathcal A$ is a Metzler matrix because
all its off-diagonal entries are non-negative~\cite{Ra11}.

A standard property of Hurwitz Metzler
matrices therefore guarantees the existence of a vector
\[
a=
\begin{pmatrix}
a_1&a_2&a_3&a_4
\end{pmatrix}^{\mathsf T}
\gg0
\]
such that
\[
a^{\mathsf T}\mathcal A\ll0,
\]
see Rantzer~\cite[Prop.~1]{Ra11}. Therefore, there exists a vector
\[
q=
\begin{pmatrix}
q_1&q_2&q_3&q_4
\end{pmatrix}^{\mathsf T}
\gg0
\]
such that
\[
a^{\mathsf T}(\mathcal F-\mathcal V)
=
-q^{\mathsf T}.
\]

Now we define a linear Lyapunov function $\mathcal{L}$ for $\psi_{\rm DFE}$ given by $\mathcal{L} = a^\intercal X$, which is more explicitly written as
\[
\mathcal L(X)
= a_1 E_c + a_2 I_c + a_3 I_b + a_4 B.
\]
Since $a\gg0$ and $X\geq0$,
\[
\mathcal L(X)\geq0,
\]
with equality if and only if
\[
X=0.
\]

Differentiating $\mathcal L$ along solutions of the system gives
\[
\frac{d\mathcal L(X)}{dt}
=
a^{\mathsf T}\dot X.
\]
Using the componentwise comparison
\[
\dot X\leq(\mathcal F-\mathcal V)X
\]
and the positivity of $a$, we obtain
\[
\frac{d\mathcal L(X)}{dt}
\leq
a^{\mathsf T}(\mathcal F-\mathcal V)X
=
-q^{\mathsf T}X.
\]
Therefore,
\[
\frac{d\mathcal L(X)}{dt}
\leq
-\left(
q_1E_c+q_2I_c+q_3I_b+q_4B
\right)
\leq0.
\]
Because $q\gg0$, equality holds only when
\[
E_c=I_c=I_b=B=0.
\]

On the infection-free set, the susceptible equations reduce to
\[
\dot S_c=\Lambda_c-\mu_{dc}S_c,
\qquad
\dot S_b=\Lambda_b-\mu_{db}S_b.
\]
Their solutions satisfy
\[
S_c(t)
=
S_c^*+
\left[S_c(0)-S_c^*\right]e^{-\mu_{dc}t},
\]
and
\[
S_b(t)
=
S_b^*+
\left[S_b(0)-S_b^*\right]e^{-\mu_{db}t}.
\]
Consequently,
\[
S_c(t)\longrightarrow S_c^*,
\qquad
S_b(t)\longrightarrow S_b^*
\qquad\text{as }t\to\infty.
\]

Thus, the largest invariant subset of
$\{\frac{d\mathcal L(X)}{dt}=0\}$ is the singleton containing the disease-free equilibrium, $\psi_{\text{DFE}}$. By LaSalle's invariance principle, the
disease-free equilibrium $\psi_{\text{DFE}}$ is globally asymptotically
stable in the biologically feasible region $\mathcal E$ whenever
$\mathcal R_0<1$.  When $\mathcal R_0>1$, the next-generation matrix result gives
\[
s(\mathcal F-\mathcal V)>0.
\]
Therefore, the linearisation at $\psi_{\mathrm{DFE}}$ has an eigenvalue
with positive real part, and the disease-free equilibrium is
unstable.}

\subsection{Global stability of the endemic equilibrium}
\myEdit{Suppose that system~\eqref{eqn:reduced} admits a unique strictly
positive endemic equilibrium
\[
\psi_{\text{EE}}
=
\left(
S_c^{**},E_c^{**},I_c^{**},
S_b^{**},I_b^{**},B^{**}
\right).
\]
At this equilibrium, the following identities hold:
\begin{equation}
\begin{aligned}
    \Lambda_c &= \beta_cS_c^{**}I_c^{**} + \beta_{bc}S_c^{**}I_b^{**}+\beta_{\rm env}S_c^{**}B^{**} + \mu_{dc}S_c^{**}, \\
    k_E E_c^{**} &= \beta_cS_c^{**}I_c^{**} +\beta_{bc}S_c^{**}I_b^{**}+\beta_{\rm env}S_c^{**}B^{**}, \\
    \sigma_cE_c^{**} &= k_cI_c^{**}, \\
    \Lambda_b &= \beta_bS_b^{**}I_b^{**} + \beta_{\rm env}S_b^{**}B^{**} + \mu_{db}S_b^{**}, \\
    k_bI_b^{**} &= \beta_bS_b^{**}I_b^{**} + \beta_{\rm env}S_b^{**}B^{**}, \\
    \mu_{\rm env}B^{**} &= \theta_cI_c^{**} + \theta_bI_b^{**},
\end{aligned}
\end{equation}
where we define
$$
k_E = \sigma_c +\mu_c+\mu_{dc}, \qquad k_c = \gamma_c + \mu_c +\mu_{dc}, \qquad k_b = \gamma_b +\mu_b +\mu_{db}.
$$
For convenience, define
$$
A = \beta_cS_c^{**}I_c^{**}, \qquad C = \beta_{bc}S_c^{**}I_b^{**}, \qquad D = \beta_{\rm env}S_c^{**}B^{**}, \qquad G = \beta_bS_b^{**}I_b^{**},
$$
$$
H = \beta_{\rm env}S_b^{**}B^{**}, \qquad M = \theta_cI_c^{**}, \qquad N = \theta_bI_b^{**}
$$
Thus, we can write $$k_E E_c^{**} = A+C+D,$$ $$k_bI_b^{**} = G+H,$$ and $$\mu_{\rm env}B^{**} = M+N.$$ We also define $$\rho = \frac{A+C+D}{\sigma_cE_c^{**}} = \frac{k_E}{\sigma_c}$$ and define the positive constants 
$$
\eta = \frac{C+D}{M}, \qquad \delta = \frac{C+\eta N}{H}.
$$

Because the endemic equilibrium and all transmission and shedding parameters are assumed strictly positive, the constants $A,C,D,G,H,M,N,\rho,\eta$, and $\delta$ are positive.

Next we consider a Volterra-type Lyapunov function~\cite{mayengo2023volterra, farman2024global} as given below:
\begin{align}
 \mathcal{L}  &= \left(S_c - S_c^{**} - S_c^{**}\ln \left(\frac{S_c}{S_c^{**}}\right) \right) +\left(E_c - E_c^{**} - E_c^{**}\ln \left(\frac{E_c}{E_c^{**}}\right) \right) +
 \rho\left(I_c - I_c^{**} - I_c^{**}\ln \left(\frac{I_c}{I_c^{**}}\right) \right) \nonumber \\
 &+\delta \left(S_b - S_b^{**} - S_b^{**}\ln \left(\frac{S_b}{S_b^{**}}\right) \right)
 +\delta \left(I_b - I_b^{**} - I_b^{**}\ln \left(\frac{I_b}{I_b^{**}}\right) \right) 
 +\eta \left(B - B^{**} - B^{**}\ln \left(\frac{B}{B^{**}}\right) \right)
\end{align}

For $r>0$
$$
r - 1 -\ln r \ge 0,
$$
with equality if and only if $r=1$. Therefore, $\mathcal L\geq0$, with equality if and only if the system is at the endemic equilibrium.

$$
x =\frac{S_c}{S_c^{**}}, \qquad e = \frac{E_c}{E_c^{**}}, \qquad i = \frac{I_c}{I_c^{**}},
$$

$$
y =\frac{S_b}{S_b^{**}}, \qquad j = \frac{I_b}{I_b^{**}}, \qquad z = \frac{B}{B^{**}}.
$$

We use the differentiation of $\mathcal{L}$ along the trajectories of the system~\eqref{eqn:reduced}, followed by application of the endemic-equilibrium identities and
collection of terms according to the equilibrium transmission and shedding flows.

Using the positive constants, we can rewrite~\eqref{eqn:reduced} as 
\begin{equation}
\begin{aligned}
    \dot{S}_c &= \mu_{dc}S_c^{**}(1 - x) + A(1-ix) +C(1-jx) + D(1-xz), \\
    \dot{E}_c &= A(ix - e) + C(jx - e) + D(xz - e), \\
    \dot{I}_c &= \sigma_cE^{**}_c(e - i), \\
    \dot{S}_b &= \mu_{db}S_b^{**}(1-y) + G(1-jy) + H(1-yz), \\
    \dot{I}_b &= Gj(y - 1) + H(yz - j), \\
    \dot{B} &= M(i - z) + N(j - z).
\end{aligned}
\end{equation}
Putting all these in $\frac{d\mathcal{L}}{dt}$ gives 
\begin{align}
\frac{d\mathcal{L}}{dt} &= \left(1 - \frac{1}{x} \right)[\mu_{dc}S_c^{**}(1 - x) + A(1-ix) +C(1-jx) + D(1-xz)]\nonumber \\
&+\left(1-\frac{1}{e} \right)[A(ix - e) + C(jx - e) + D(xz - e)]\nonumber \\
&+\rho \left(1 - \frac{1}{i} \right)[\sigma_cE^{**}_c(e - i)]+\delta \left(1 - \frac{1}{y} \right)[\mu_{db}S_b^{**}(1-y) + G(1-jy) + H(1-yz)] \nonumber \\
&+\delta \left(1 - \frac{1}{j} \right) [Gj(y - 1) + H(yz - j)] +\eta\left(1-\frac{1}{z}\right)[M(i - z) + N(j - z)].
\end{align}
Now, using the identities of $\rho, \eta$ and $\delta$
\begin{enumerate}
    \item[(i)] First look at the $\mu_{dc}S_c^{**}$ contribution: $\mu_{dc}S_c^{**}\left(1 - \frac{1}{x} \right)(1-x)$. Now $\left(1 - \frac{1}{x} \right)(1-x) = 2 - x -\frac{1}{x}$.
    \item[(ii)] Similarly, collecting the contribution of $\delta \mu_{db}S_b^{**}$ we get $2 - y -\frac{1}{y}$.
    \item[(iii)] Collecting the coefficients of $A$ we get $\left(1 - \frac{1}{x} \right)(1-ix) + \left(1-\frac{1}{e} \right)(ix - e) + \left(1 - \frac{1}{i} \right)(e-i)$ which on expanding and adding becomes $3 - \frac{1}{x} -\frac{ix}{e} - \frac{e}{i}$.
    \item[(iv)] Similarly collecting the coefficients of $G$ gives us $\delta \left(2 - y -\frac{1}{y} \right)$.
    \item[(v)] Collecting the coefficients of $D$ gives us $\left(1 - \frac{1}{x}\right)(1-xz) + \left(1-\frac{1}{e} \right)(xz - e) + \left(1 - \frac{1}{i} \right)(e - i) + \left(1-\frac{1}{z}\right) (i - z)$. Expanding and adding this becomes $\left(4 - \frac{1}{x} - \frac{xz}{e} -\frac{e}{i} - \frac{i}{z} \right)$.
    \item[(vi)] Collecting the coefficients of $C$ gives us the longest term which on expanding and summing similarly gives us $\left(6 - \frac{1}{x} -\frac{xj}{e} - \frac{e}{i} -\frac{i}{z} -\frac{1}{y} - \frac{yz}{j} \right)$.
    \item[(vii)] Finally, we collect the coefficients of $N$, which is $\eta \left[\left( 1-\frac{1}{y}\right)(1-yz) + \left(1-\frac{1}{j} \right)(yz-j) + \left(1-\frac{1}{z} \right)(j-z) \right]$. On expanding and summing this, we get $\eta\left(3 - \frac{1}{y} -\frac{yz}{j} - \frac{j}{z} \right)$.
\end{enumerate}
Thus, putting all these together, we get
\begin{align}
  \frac{d\mathcal{L}}{dt} &= \mu_{dc}S_c^{**}\left(2-x-\frac{1}{x}\right) + \delta \mu_{db}S_b^{**}\left(2-y-\frac{1}{y}\right) + A\left(3-\frac{1}{x} -\frac{ix}{e} - \frac{e}{i} \right) \nonumber \\
  &+ \delta G\left(2 - y -\frac{1}{y} \right) + D\left(4 - \frac{1}{x} - \frac{xz}{e} -\frac{e}{i} - \frac{i}{z} \right) + C\left(6 - \frac{1}{x} -\frac{xj}{e} - \frac{e}{i} -\frac{i}{z} -\frac{1}{y} - \frac{yz}{j} \right) \nonumber \\
  &+ \eta N \left(3 - \frac{1}{y} -\frac{yz}{j} - \frac{j}{z} \right).
\end{align}


Each term on the right-hand side of $\frac{d\mathcal{L}}{dt}$ is non-positive. The first, second, and fourth terms follow from $r+r^{-1}\geq2$ for $r>0$, while the remaining terms follow from the arithmetic--geometric mean inequality. For any positive numbers $a_1, \ldots, a_n$
$$
\frac{a_1+ \ldots +a_n}{n} \ge (a_1a_2\ldots a_n)^{1/n}.
$$
For example 
$$
\frac{\frac{1}{x} +\frac{ix}{e} + \frac{e}{i}}{3} \ge \left(\frac{1}{x} \times \frac{ix}{e}  \times \frac{e}{i} \right)^{1/3}=1.
$$

The products of the ratios in each of the other terms are also equal to one. Consequently, $\frac{d\mathcal{L}}{dt} \leq0$ throughout the interior of the biologically feasible region.

Equality requires $x=y=1$ and $e=i=j=z$. On this set, the susceptible cattle equation gives
\[
\dot S_c=(A+C+D)(1-i).
\]
Because $A+C+D>0$, invariance requires $i=1$. Hence,
\[
x=e=i=y=j=z=1,
\]
and the largest invariant subset of $\frac{d\mathcal{L}}{dt} =0$ is the singleton containing the endemic equilibrium. Therefore, by LaSalle's invariance principle, the unique strictly positive endemic equilibrium is globally asymptotically stable in the interior of the biologically feasible region.

}

\end{document}